\documentclass{bmcart}

\usepackage{amsthm,amsmath}
\usepackage[utf8]{inputenc} 
\usepackage{graphicx}   

\startlocaldefs
\endlocaldefs

\begin{document}

\begin{frontmatter}

\begin{fmbox}
\dochead{Research}


\title{Engineering the microwave to infrared noise photon flux for superconducting quantum systems}


\author[
  addressref={aff1},                   
  corref={aff1},                       
  email={sergey.danilin@glasgow.ac.uk}   
]{\inits{S.}\fnm{Sergey} \snm{Danilin}}
\author[
  addressref={aff1},
]{\inits{J.}\fnm{Jo\~ao} \snm{Barbosa}}
\author[
addressref={aff1},
]{\inits{M.}\fnm{Michael} \snm{Farage}}
\author[
addressref={aff1},
]{\inits{Z.}\fnm{Zimo} \snm{Zhao}}
\author[
addressref={aff2},
]{\inits{X.}\fnm{Xiaobang} \snm{Shang}}
\author[
addressref={aff2},
]{\inits{J.}\fnm{Jonathan} \snm{Burnett}}
\author[
addressref={aff2},
]{\inits{N.}\fnm{Nick} \snm{Ridler}}
\author[
addressref={aff1},
]{\inits{C.}\fnm{Chong} \snm{Li}}
\author[
addressref={aff1},
]{\inits{M.}\fnm{Martin} \snm{Weides}}
	

\address[id=aff1]{
  \orgdiv{James Watt School of Engineering},             
  \orgname{University of Glasgow},          
  \city{Glasgow} 
  \postcode{G12 8QQ},                             
  \cny{UK}                                    
}
\address[id=aff2]{%
  \orgdiv{},
  \orgname{National Physical Laboratory},
  \street{Hampton Road},
  \city{Teddington}
  \postcode{TW11 0LW}
  \cny{UK}
}



\end{fmbox}


\begin{abstractbox}

\begin{abstract} 
Electromagnetic filtering is essential for the coherent control, operation and readout of superconducting quantum circuits at milliKelvin temperatures. The suppression of spurious modes around transition frequencies of a few GHz is well understood and mainly achieved by on-chip and package considerations. Noise photons of higher frequencies -- beyond the pair-breaking energies -- cause decoherence and require spectral engineering before reaching the packaged quantum chip. The external wires that pass into the refrigerator and go down to the quantum circuit provide a direct path for these photons. This article contains quantitative analysis and experimental data for the noise photon flux through coaxial, filtered wiring. The attenuation of the coaxial cable at room temperature and the noise photon flux estimates for typical wiring configurations are provided. Compact cryogenic microwave low-pass filters with CR-110 and Esorb-230 absorptive dielectric fillings are presented along with experimental data at room and cryogenic temperatures up to 70 GHz. Filter cut-off frequencies between 1 to 10 GHz are set by the filter length, and the roll-off is material dependent. The relative dielectric permittivity and magnetic permeability for the Esorb-230 material in the pair-breaking frequency range of 75 to 110 GHz are measured, and the filter properties in this frequency range are calculated. The estimated dramatic suppression of the noise photon flux due to the filter proves its usefulness for experiments with superconducting quantum systems.
\end{abstract}


\begin{keyword}
\kwd{coaxial cable attenuation}
\kwd{noise photon flux}
\kwd{material electromagnetic properties}
\end{keyword}


\end{abstractbox}
%

\end{frontmatter}



\section{\label{sec:introduction}{Introduction}}

Superconducting quantum circuits are a mature and salient experimental platform for the development of quantum technologies~\cite{QuantuGuide2019_Oliver}. They are at the core of technological transition to a so-called Noisy Intermediate-Scale Quantum (NISQ) level~\cite{NISQ2018_Preskill}, where they are used for the construction of multi-qubit processors for quantum computation~\cite{Supremacy2019} and for the creation of structures to work as quantum simulators of other physical systems that are hard to study in a laboratory~\cite{On_chip_simul_2012_Koch,2D_Bose_Hubbard_Tahan}. They also find applications in the sensing of amplitude, frequency~\cite{Local_sensing,Amplitude_and_frequency_sensing} and power~\cite{Quantum_sensor_of_power} of microwave signals and in quantum metrology~\cite{Quantum-enhanced_magnetometry,Magnetic_field_detection}. For all of these tasks a quantum circuit needs to be well protected from external sources of decoherence, and precise control of the quantum state of the circuit and fast readout are required. Rapid control is performed by quickly changing signals delivered to the quantum structure via coaxial wiring lines. In addition, these coaxial lines (for drive, flux control and readout~\cite{100Q_wiring_ETH_Krinner}) bring electromagnetic noise to the structure and create additional channels of decoherence. The spectrum of noise can cover a wide range of frequencies, but control and readout are implemented within quite narrow frequency bands. Therefore, microwave attenuation, filtering and shielding are essential techniques widely used in experiments with superconducting quantum circuits. In these experiments, a superconducting circuit is placed in a cryogenic refrigerator, at a temperature of $\sim 10\ {\rm mK}$, where it is shielded from stray magnetic fields and thermal radiation. Coaxial wiring for control and readout is thermally anchored at all temperature stages and attenuated and filtered at some of them. After the circuit is interrogated with readout signals, these signals are amplified and also filtered.

Radiation impinging on the circuit with frequencies outside the frequency bands
of the control and readout signals is detrimental to its quantum state and needs to be filtered. Quite often low-pass filters with GHz cut-off frequencies start to transmit again at higher frequencies close to the infrared range~\cite{www.minicircuits.com}.  Radiation with frequency $\nu$, having energy $h\nu>2\Delta$, where $\Delta$ is the superconducting energy gap, and $h$ is the Planck's constant, breaks Cooper pairs, and hereby generates, in the bulk of the circuit electrodes, quasiparticles detrimental to the coherence of quantum states. Moreover, the mechanism of qubit decoherence associated with the photon-assisted electron tunnelling through a Josephson junction has also been identified~\cite{photon_assisted_decoherence}. For aluminium, with superconducting energy gap $\Delta\simeq 170\ \mu eV$ for film thickness $\sim 100\ {\rm nm}$~\cite{Douglass_Al_energy_gap}, this corresponds to frequencies $\nu>82$ GHz. In addition, it was shown that thermal radiation can generate fluctuations in the residual photon number and dephase superconducting qubits due to the ac Stark effect~\cite{Yan_thermal_photon_dephasing}. Also, the nonthermal populations of higher resonator modes are important for qubit dephasing~\cite{higher_mode_dephasing}. 

There are two paths for the unwanted radiation to reach the circuit: direct impingement from higher temperature stages (free-space photons), and through TEM, TE and TM modes propagating in microwave coaxial wiring. It has previously been demonstrated~\cite{Infrared_shielding_UCSB_Barends,Radiation_protection_IBM_Corcoles} that shielding of superconducting quantum circuits from infrared radiation is efficient to suppress the radiation flow through the first path. This infrared shielding improves quality factors of resonators and relaxation times of qubits. Control and readout signals reach the structure through the second path, making the attenuation and filtering conditions more stringent. Here cryogenic attenuators are commonly used to lower the signal levels and reduce the number of thermal photons reaching the structure~\cite{100Q_wiring_ETH_Krinner, Yeh_microwave_attenuators, Yeh_heatsinks_for_microwave_attenuators}. 

In our work, we provide a compact review of existing microwave filters (Sec.~\ref{sec:cryogenic_filters}) before calculating the attenuation of different microwave modes and estimating the flow of noise photons in a standard coaxial wire (Sec.~\ref{sec:noise_photons}). We then (Sec.~\ref{sec:material_design}) demonstrate the design and manufacture of low-pass filters suitable for use in a cryogenic environment. In Sec.~\ref{sec:s_parameters}, the microwave properties of filters are characterised at room temperature and $\sim 3$~K up to $70$ GHz. Sec.~\ref{sec:Esorb_electromagnetic_parameters} provides the results of measurements of Esorb-230 material electromagnetic properties in the 75 to 110 GHz frequency range. Finally, the reduction of residual noise photon flux when the microwave filters are used is estimated in Sec.~\ref{sec:residual_photons}, which demonstrates their utility in experiments with superconducting quantum circuits and that Esorb-230 is suitable for the fabrication of infrared shields.

\section{\label{sec:cryogenic_filters}{Overview on cryogenic microwave low-pass filters}}

Filters are used to limit the allowed frequency pass-band. In the following, we provide a short overview of cryogenic microwave low-pass filters that have been developed to date. The most common and widely used type is metal powder filter, of which there are many variations: made of copper and stainless steel powder~\cite{Copper_powder_Martinis,Powder_Fukushima}, based on silver epoxy~\cite{Silver-epoxy_Scheller}, 50$\Omega$-matched bronze and stainless steel powder~\cite{50_matched_bronze_powder}. There are also varieties with stripline embedded in magnetically-loaded Eccosorb dielectric~\cite{50_matched_stripline_Eccosorb}, with built-in capacitive shunts to lower the cut-off frequency~\cite{SS_powder_Lukashenko}, and ones based on printed circuit boards (PCB) embedded in metal powder~\cite{PCB_metal_powder}. Other types are micro-fabricated miniature filters based on lossy coplanar transmission lines~\cite{Vion_meander_line}, and on-chip filters for the millimetre frequency range, comprising arrays of SQUIDs or resistive capacitively shunted transmission lines~\cite{JAF_RCF_Lehtinen}. Thin Thermocoax and stainless steel cables have filtering properties themselves and were tested as microwave cryogenic filters for experiments with quantum circuits~\cite{thermocoax_Zorin,Trancredi_SS_coax}.  Different types of microwave filters are compared in reviews Ref.~\cite{Bladh_filter_comparison} and Ref.~\cite{Thalmann_filter_comparison}. The latter also introduces a transmission line ferrite compound filter. Most of these filters are tested experimentally at sub-GHz frequencies up to $20$ GHz and only the metal powder filters reported in Ref.~\cite{Bladh_filter_comparison} were measured up to $50$ GHz, which is still lower than the frequency corresponding to the aluminium superconducting energy gap. Both the lack of literature on filter transmission above $20$ GHz and the superconducting circuit's sensitivity to stray radiation, particularly at these frequencies, are addressed in the following sections.


\section{\label{sec:noise_photons}{Estimation of noise photon flux}}

\begin{figure}[h!]
	\includegraphics[width=0.96\textwidth]{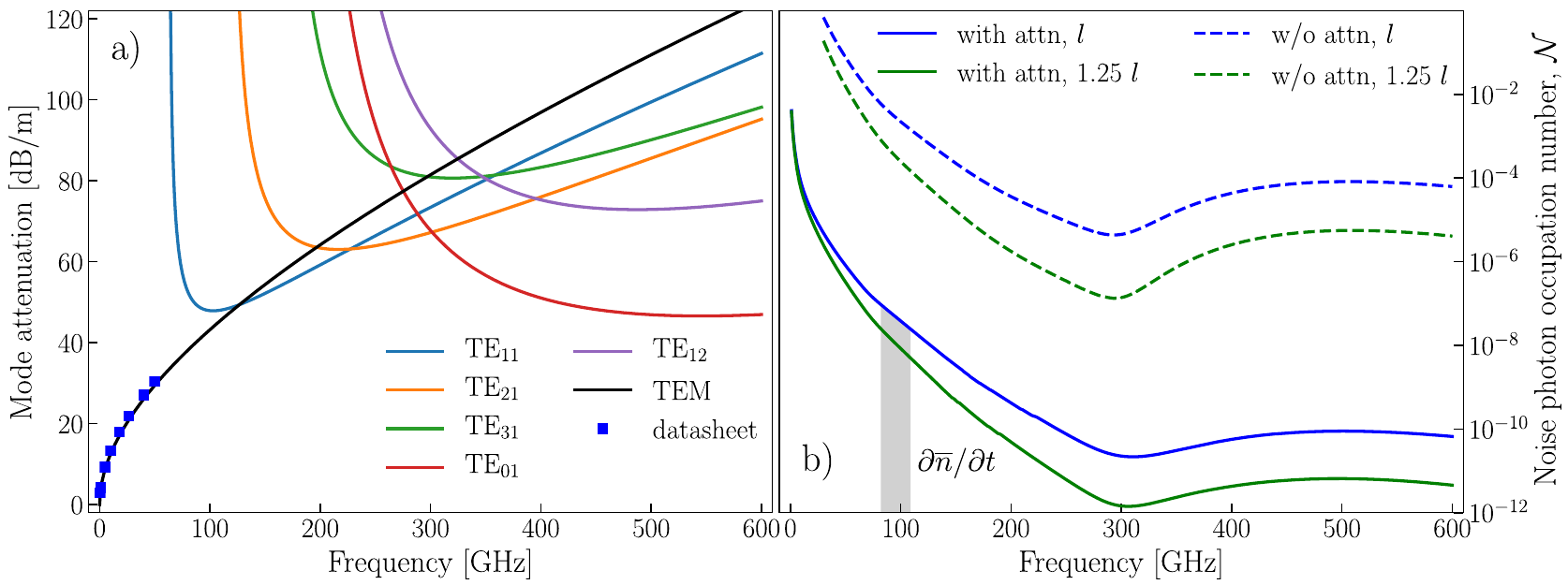}
	\caption{(a) Room temperature attenuation caused by conductor and dielectric losses in UT086SS-SS cable for the six modes with the lowest attenuation. All TM modes have higher attenuation. Blue squares show the data from Ref.~\cite{micro_coax} for the TEM mode. (b) Estimate of the noise photon occupation number at the mixing chamber stage and for a single coaxial cable without filtering computed as a sum of contributions caused by each of the modes: for attenuators (assuming a constant attenuation over the entire frequency range) and two different lengths of coaxial line -- two lower lines, without attenuators -- two higher lines. Here an ideal thermalization of coaxial line and attenuators is assumed. Shaded area denotes the average number of noise photons ($\partial\overline{n}/\partial t = \int\mathcal Nd\nu$) exiting the coaxial line per second at the mixing chamber in the 82 to 110 GHz frequency range. See Appendix 1 and 2 for the details of calculation.}	
	\label{motivation_figure}
\end{figure}

Fig.~\ref{motivation_figure}(a) shows room temperature attenuation per unit length $\alpha(\omega)$[dB/m] of the five microwave TE modes with the lowest attenuation and a TEM mode for UT086SS-SS stainless steel coaxial cable with a PTFE dielectric, including conductor and dielectric losses (see Appendix 2). A range of up to 600 GHz is chosen to include the global minimum ($\sim 47\ {\rm dB/m}$) 
of attenuation for TE modes at $\sim 545\ {\rm GHz}$, where the number of noise photons transmitted through the line reaches a local maximum and then decreases due to the increased attenuation for higher frequencies.
TM modes all have much higher attenuation, with a minimum of $\sim 100\ {\rm dB/m}$, and cannot noticeably contribute to the transmission of radiation. Higher order TE modes have attenuations $\sim 50\ {\rm dB}$ higher than the minimal attenuation in this frequency range and their contribution to the transmission is negligible (less than $0.5\%$ of the total transmission). 

We have developed a model required to compute the residual photon population (see Appendix 1). The model includes the frequency-dependent attenuation, and it is a significant improvement over past work~\cite{100Q_wiring_ETH_Krinner}. We estimate the noise photon occupation numbers $\mathcal{N}(x,\omega)=\partial^2{\overline{n}}/\partial\nu\partial t$ -- the average number of photons passing through a cross-section of the cable per unit bandwidth in a second -- at the mixing chamber stage (Fig.~\ref{motivation_figure}(b)) by solving the equation

\begin{equation}
\frac{\partial \mathcal{N}(x,\omega)}{\partial x}=\frac{\alpha(\omega)\ln{10}}{10}\bigg(n_{BE}\big(\omega,T(x)\big)-\mathcal{N}\big(x,\omega\big)\bigg)
\label{occupation_number}
\end{equation}
consecutively for the coaxial line sections connecting the temperature stages of the refrigerator. Here, $\bar{n}$ is the average number of photons passed through cross-section of the cable, $x$ denotes the position along the coaxial line from the point at room temperature where the cable enters the refrigerator, and $n_{BE}(\omega,T(x))$ is the Bose-Einstein distribution at frequency $\omega$ and temperature $T(x)$. Temperatures of refrigerator stages, lengths of coaxial cables between the stages, and the arrangement of attenuators at the stages are given in Appendix 1. At those temperature stages where attenuators $a$[dB] are placed, the function $\mathcal{N}(x,\omega)$ has abrupt changes described by the equation 
\begin{equation}
\mathcal{N}(x,\omega)_{\textrm{out}}=\frac{\mathcal{N}(x,\omega)_{\textrm{in}}}{A}+\frac{A-1}{A}n_{BE}\big(\omega,T(x)\big),
\label{attenuator_occupation}
\end{equation}
where $A=10^{a[\textrm{dB}]/10}$. Attenuator scattering properties above $18\ {\rm GHz}$ are not provided by the manufacturers. Thus, we provide two estimates for i) the case when the attenuators work up to 600 GHz as well as they do up to $18\ {\rm GHz}$, and ii) the case when the attenuators do not work at all and we can consider only cable attenuation. The average flow of noise photons in the frequency range from 82 GHz up to 110 GHz (all above the Cooper pair breaking energy for aluminum) without the use of infrared filters accounts for about $\partial\overline{n}/\partial t\sim1420$ photons a second with attenuators ($\sim 85$ million photons a second without attenuators). Extending the frequency range up to 600 GHz enlarges the noise photon flow by 35\% of the value for 82-110 GHz range (by 56\% without attenuators). Increasing the length of coaxial cables between each pair of refrigerator stages by 25\% of their initial length lowers the average flow of noise photons to $\sim 335$ ($\sim 10$ million). This modest change in length has a significant impact of more than a factor 4 in the noise photon flux. This elucidates the strong dependence of the pair breaking photon flux without additional infrared filtering on the individual coaxial wiring within the cryostat. Lines in Fig.~\ref{motivation_figure}(b) represent the lower bounds on the noise photon occupation numbers as the modes' attenuation is reduced at lower temperatures and this temperature dependence is not taken into account. As a note, thinner coaxial cables, such as UT047SS-SS and UT034SS-SS, have higher attenuation per meter which leads to a reduction of the noise photon flux, see Fig.~\ref{fig7} and Appendix 2.

\section{\label{sec:material_design}{Filter Materials and Design}}
Our microwave filters consist of two microwave connectors in a hollow copper block with their PTFE removed. The centre conductor ends are soldered together, and the filter material is cast around the centre pins. We use a design of enclosure blocks similar to that of Ref.~\cite{Thesis_UCSB_Fang} and fill them with two absorptive materials: CR-110~\cite{www.laird.com} and Esorb-230~\cite{Esorb230spec}. Both materials are commercially available and easy to handle.
These materials are magnetically-loaded epoxy absorbers which consist of a mixture of a low dielectric loss matrix with micrometer-scale magnetic particles, that produce a high loss tangent~\cite{zivkovic2011characterization}. Together with a curing agent, they form a rigid material that can be cast to fit around the central conductor and act as a filter device. Absorption properties of these two materials are distinct not only due to the choice of dielectric material used, but more importantly because of the difference in density of magnetic particles in each mixture.

\begin{figure}[h!]
	\includegraphics[width=0.96\textwidth]{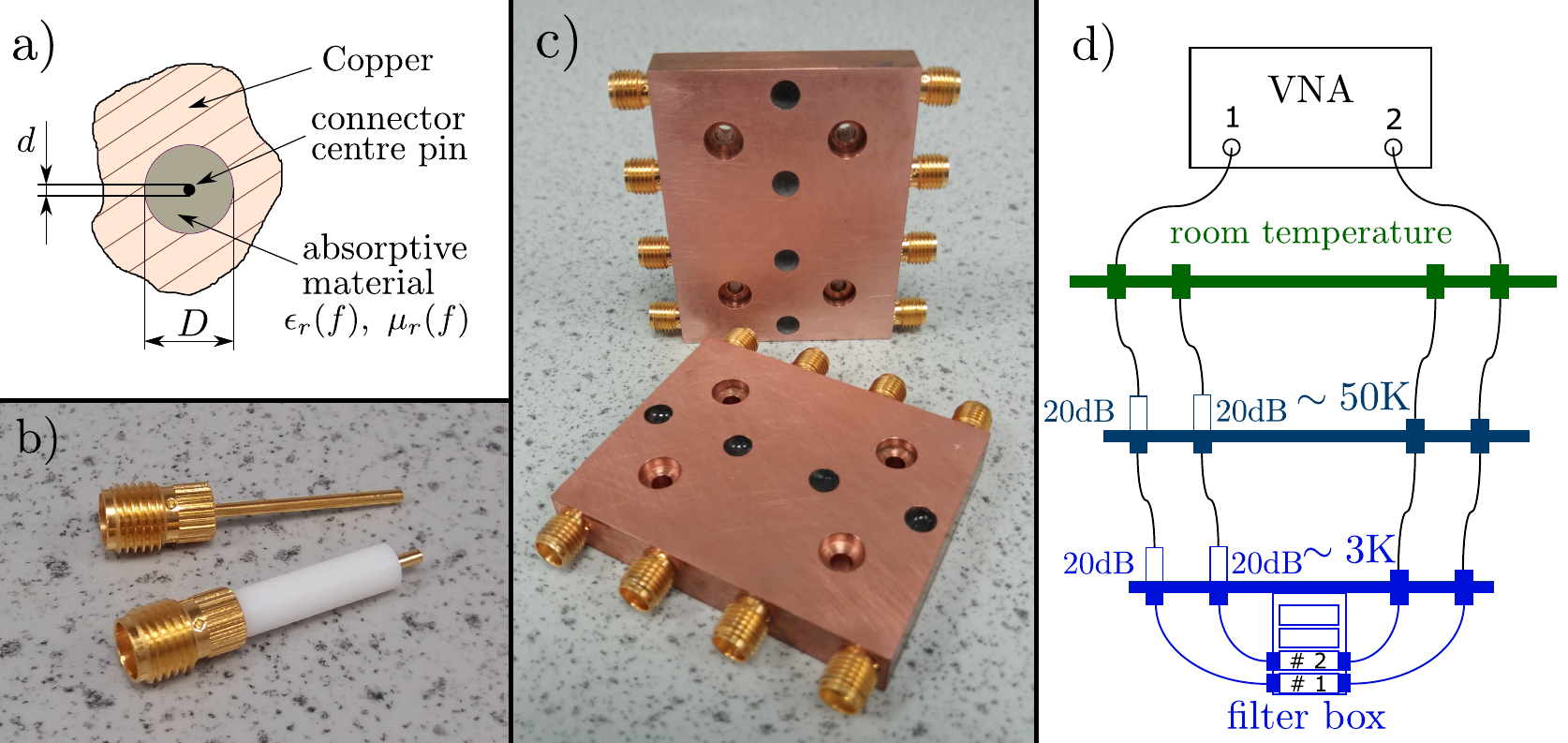}
	\caption{a) Schematic view of a cross-section of the filter. b) Real image of SMA connectors used for the fabrication of filters. c) Real image of microwave filters tested in the work. d) Diagram of the setup used in cryogenic measurements of scattering parameters.}
	\label{fig:setup}
\end{figure}

The filter is made by drilling a channel of diameter $D$ in a copper block (Fig.~\ref{fig:setup}(a),(c)). Two SMA connectors (Johnson 142-1721-051, knurl mount, Fig.~\ref{fig:setup}(b)) are plugged in the channel from both sides so that the centre conductors meet in the middle of the channel and can be soldered together. The dielectric layer (PTFE) around the centre conductor is removed before the installation. 

Finally, the volume between the centre conductor and the walls of the channel is filled with either CR-110 or Esorb-230 absorptive material through the hole in the copper enclosure in the middle of the channel. The same hole was used before to solder the ends of centre conductors.

Given the coaxial geometry of the filter, the characteristic impedance can be computed as~\cite{Thesis_UCSB_Fang}
\begin{equation}
Z(f)=\frac{Z_{\rm{vac}}\ln{\left(D/d\right)}}{2\pi}\sqrt{\frac{\mu_r(f)}{\epsilon_r(f)}}.
\label{eq:impedance}
\end{equation}
Here, $Z_{\rm{vac}}=\sqrt{\mu_0/\epsilon_0}\simeq 377\Omega$ is the impedance of free space, $D$ is the diameter of the channel in copper, $d$ is the diameter of centre pin of the connector, and $\mu_r(f)=\mu'(f)-j\mu''(f)$ and $\epsilon_r(f)=\epsilon'(f)-j\epsilon''(f)$ are the relative magnetic permeability and electric permittivity of the absorptive material. These material constants, and hence the optimal diameter $D$, depend on the frequency $f$~\cite{www.laird.com}. Using the centre pin diameter, $d=1.27$ mm, of the connectors, we find the outer diameter, $D^*=5.3$ mm, which minimises reflection $20\log_{10}{\left|(Z-50\Omega)/(Z+50\Omega)\right|}$ averaged in 1 to 18 GHz frequency range. The actual diameter $D=5.1$ mm used in the manufacture, and given by the size of the knurled part of the connectors, is very close to the optimal value $D^*$. It gives an average reflection of $-31$ dB in the same frequency range with a maximum impedance deviation from $50~\Omega$ by $11~\Omega$ at 1 GHz.

\section{\label{sec:s_parameters}{Scattering parameters at room and cryogenic temperatures}}

All filters were initially characterised at room temperature using an Agilent Technologies E8361A vector network analyzer (VNA) to measure scattering parameters up to 70 GHz. Since the SMA connectors that are fitted to these filters are only specified to 18 GHz, measurements made above this frequency should be treated with care. This is because there is a likelihood that modes other than TEM modes will also be propagating through the filter. Afterwards, the filters were tested at cryogenic temperatures inside a 3K refrigerator (see Fig.~\ref{fig:setup}(d)). A Rohde \& Schwarz ZVA 40 VNA (measuring up to 43 GHz) was used for cryogenic measurements, and two 20 dB attenuators were placed at 50K and 4K temperature stages on the input line of the setup between VNA port 1 and the filter box. In both cases, we measured "through" connections to later account for the attenuation of the wiring.

\begin{figure}[h!]
	\includegraphics[width=0.96\textwidth]{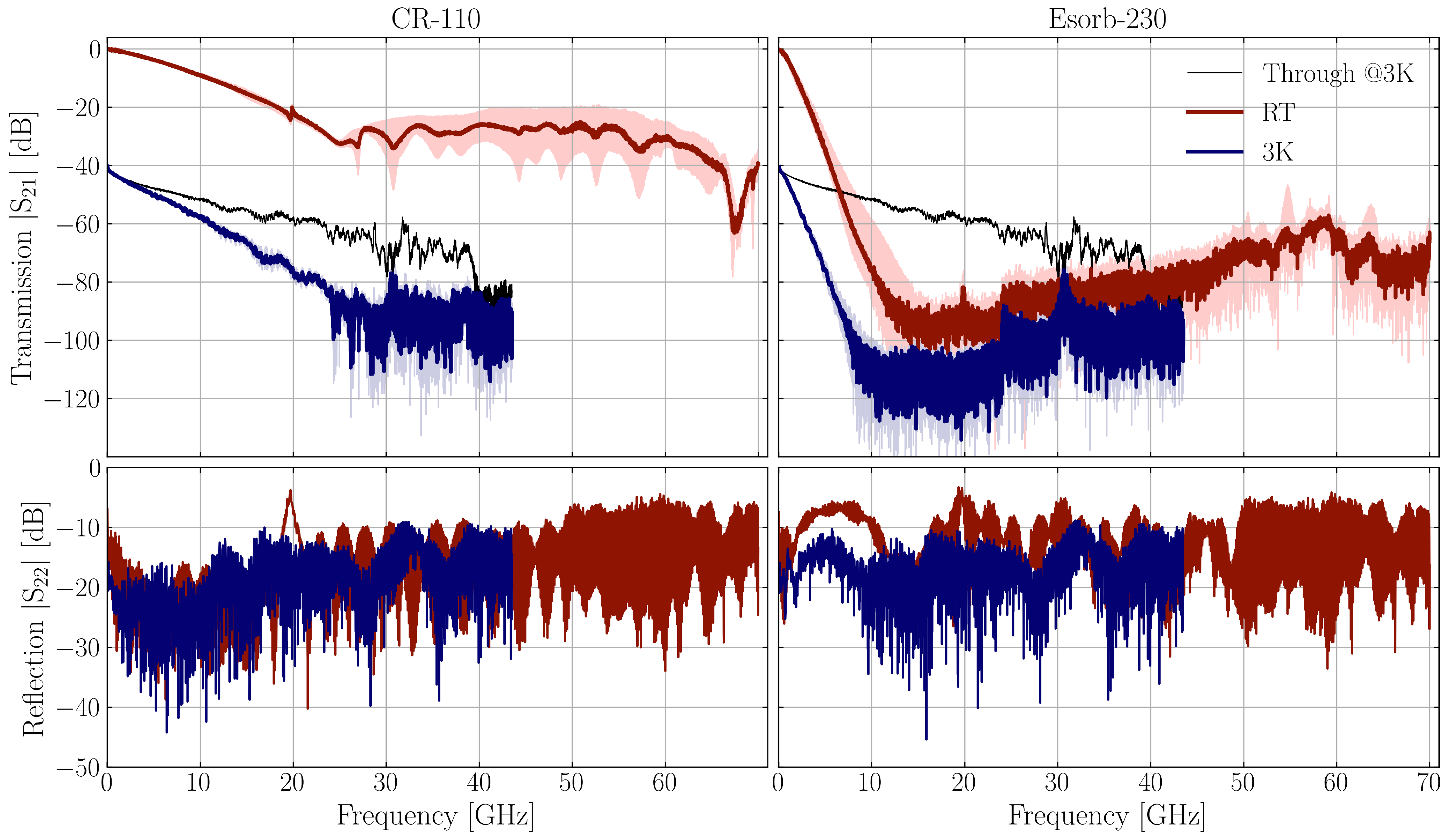}
	\caption{Reflection $\vert S_{22}\vert$ and Transmission $\vert S_{21}\vert$ probed for CR-110 and Esorb-230 filters. Characteristics were obtained for room temperature and $\sim3$ K (red and blue curves, respectively), where, for the latter, an extra 40 dB of attenuation was added to the ingoing signal. Solid lines represent the average values for 4 filters measured at room temperature and 2 filters measured cryogenically. The RT data is already corrected by the calibration with 'through' connection. The lightly shaded areas represent the biggest deviation from the average among 4 or 2 filters demonstrating the reproducibility of manufacturing.}
	\label{fig:S_parameters}
\end{figure}

Reflection $S_{22}$ and transmission $S_{21}$ were probed and the resulting data was processed to exclude the frequency-dependent behaviour of the wiring. Results are shown in Fig.~\ref{fig:S_parameters} for CR-110 and Esorb-230 filters at both temperatures. Since each box contains four filters (only two were probed at 3 K), we averaged their spectra, and the lightly shaded regions in the figure correspond to the maximum deviation from this average. Both materials are characterized by a different -3dB point and roll-off slope in transmission. For CR-110, the transmitted signal amplitude decreases slower than for Esorb-230 and reaches about -35 dB at 25 GHz.  For Esorb-230, the amplitude reaches the noise floor of the instrument at $\sim$15 GHz for room temperature measurement, and the filter transmission is covered by the noise floor for higher frequencies, see the datasheet~\cite{VNA67}. The difference in noise floor is a consequence of using two distinct VNAs. The small peak seen for CR-110 at room temperature around 20 GHz is a feature caused by a small resonance of the cable used.

For reflection data, there is a slightly higher amplitude for Esorb-230 which is a result of designing the filters using the specifications of CR-110 (data for $\epsilon_r(f)$ and $\mu_r(f)$ was not available for the former). Impedance of Esorb-230 filters was not matched to 50$\Omega$, and the signal reflection was not minimized at the filter design stage for these filters. The impedance matching of the filters with this material requires further experimental studies.

The difference in -3 dB point is even more evident when we zoom in at low frequencies, as seen in Fig. 4. The transmission data indicates that the -3 dB points shift to higher frequencies as the filters are cooled down (blue curves). The shift is approximately 50\% for CR-110 and 30\% for Esorb-230 relative to the room temperature values and can be accounted for by an increase in conductivity of metallic particles, as is usually the case for metal powder filters~\cite{Powder_Fukushima,50_matched_bronze_powder}. The  characteristics of filters are summarised in Tables \ref{table_2} and \ref{table_1}.

\begin{figure}[h!]
	\begin{minipage}[t]{0.47\textwidth}
		\vspace{0pt}
		\caption{Low frequency behaviour of filters.\newline -3 dB cut-off frequencies are distinct for both materials and increase when filters are cooled down to 3K.}
	\end{minipage}
	\begin{minipage}[t]{0.47\textwidth}
		\vspace{1mm}
		\includegraphics[width=\textwidth]{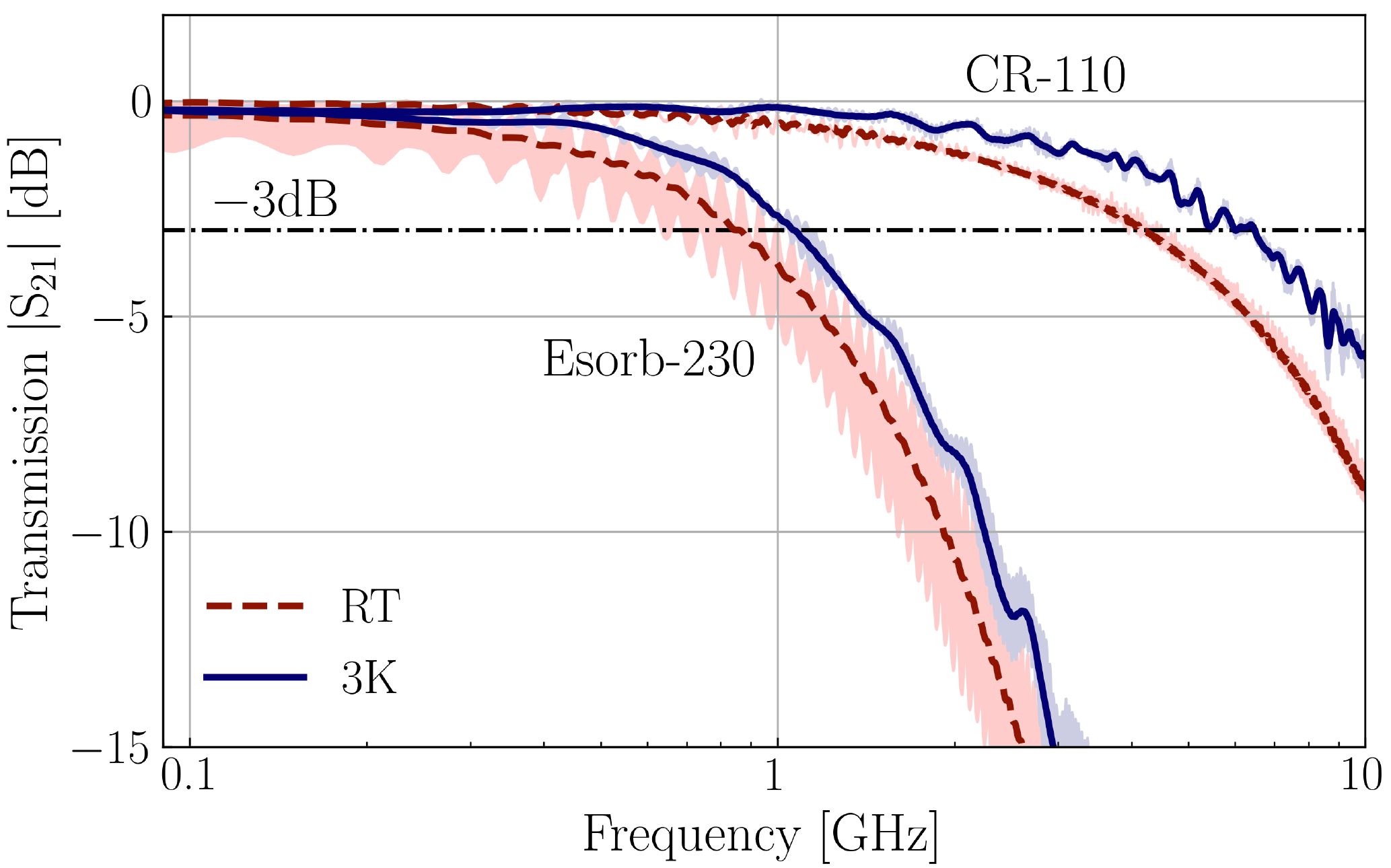}
	\end{minipage}
\label{fig:3db}
\end{figure}

\begin{table}[h!]
	\caption{Attenuation per unit length for CR-110 and Esorb-230 filters.}
	\begin{tabular}{l|c|c|c|c|c|c|c|c|}
		\hline\hline
		& \multicolumn{8}{c}{Attenuation [dB/cm]}\\
		\cline{2-9}
		& \multicolumn{2}{c}{1 GHz} & \multicolumn{2}{|c}{3 GHz} & \multicolumn{2}{|c}{5 GHz}& \multicolumn{2}{|c|}{8.6 GHz}\\
		& \multicolumn{1}{c}{RT} & 3 K & \multicolumn{1}{c}{RT} & 3 K & \multicolumn{1}{c}{RT} & 3 K & \multicolumn{1}{c}{RT} & \multicolumn{1}{c|}{3 K} \\\hline
		CR-110 & 0.155 & 0.045 & 0.550 & 0.337 & 1.034 & 0.662 &  2.070 & 1.553  \\
		CR-110 [datasheet] & 0.09 & - & 0.26 & - & - & - & 2.0 & - \\[1ex]
		Esorb-230 & 1.053 & 0.740 & 4.980 & 4.269 &  9.623 & 7.772 & 18.433 & 14.611 \\
		\hline\hline
	\end{tabular}
	\label{table_2}
\end{table}

\begin{table}[h!]
	\caption{-3 dB frequency points of filters (CR-110 and Esorb-230) and the filters attenuation at 5 GHz for room and cryogenic temperature. The shifts of -3 dB frequency points are given in \% relative to the room temperature values.}
	\begin{tabular}{l|ccc|cc}
		\hline\hline
		& \multicolumn{3}{c|}{-3dB point [GHz]} & \multicolumn{2}{c}{Atten. @ 5GHz [dB]} \\ 
		\hline
		& RT & 3 K & $\delta f_{-3\rm dB}$ & RT & 3 K \\
		CR-110 & 4.23 & 6.30 & 48.9 \% & 3.7  & 2.4   \\
		Esorb-230 & 0.83 & 1.07 & 28.9 \% & 34.6 &  28.0  \\
		\hline\hline
	\end{tabular}
	\label{table_1}
\end{table}

\section{\label{sec:Esorb_electromagnetic_parameters}{Esorb-230 characteristics in Extremely High Frequency band}}

To study absorption properties of the Esorb-230 material used for the filters at frequencies above the superconducting gap of Aluminium, we measured $S_{11}$ and $S_{21}$ parameters of waveguide sections filled with the material. WR10 rectangular waveguides $2.54\ \rm{mm}\ \times\ 1.27\ \rm{mm}$ were used in the measurements, see Fig. 5(a). Two sections of waveguides with $2.0\ \rm{mm}$ and $2.7\ \rm{mm}$ thicknesses were tested in the 75 to 110 GHz frequency range. Results of the S-parameter measurements are shown in Fig. 5(b).

\begin{figure}[h!]
	\begin{minipage}[t]{0.47\textwidth}
		\vspace{0mm}
		\caption{a) Rectangular WR10 2.54 mm $\times$ 1.27 mm waveguide sections filled with the Esorb-230 material. The spacers are used for the measurements of the $S_{11}$ and $S_{21}$ parameters in the 75 to 110 GHz frequency range. These are then used to determine the electromagnetic properties ($\epsilon_r, \mu_r$) of the material. b) Amplitude and phase of scattering parameters measured for both waveguide spacers. The solid black lines show the results of HFSS simulation with electromagnetic parameters extracted by the NRW method.}
	\end{minipage}
	\begin{minipage}[t]{0.47\textwidth}
		\vspace{1mm}
		\includegraphics[width=\textwidth]{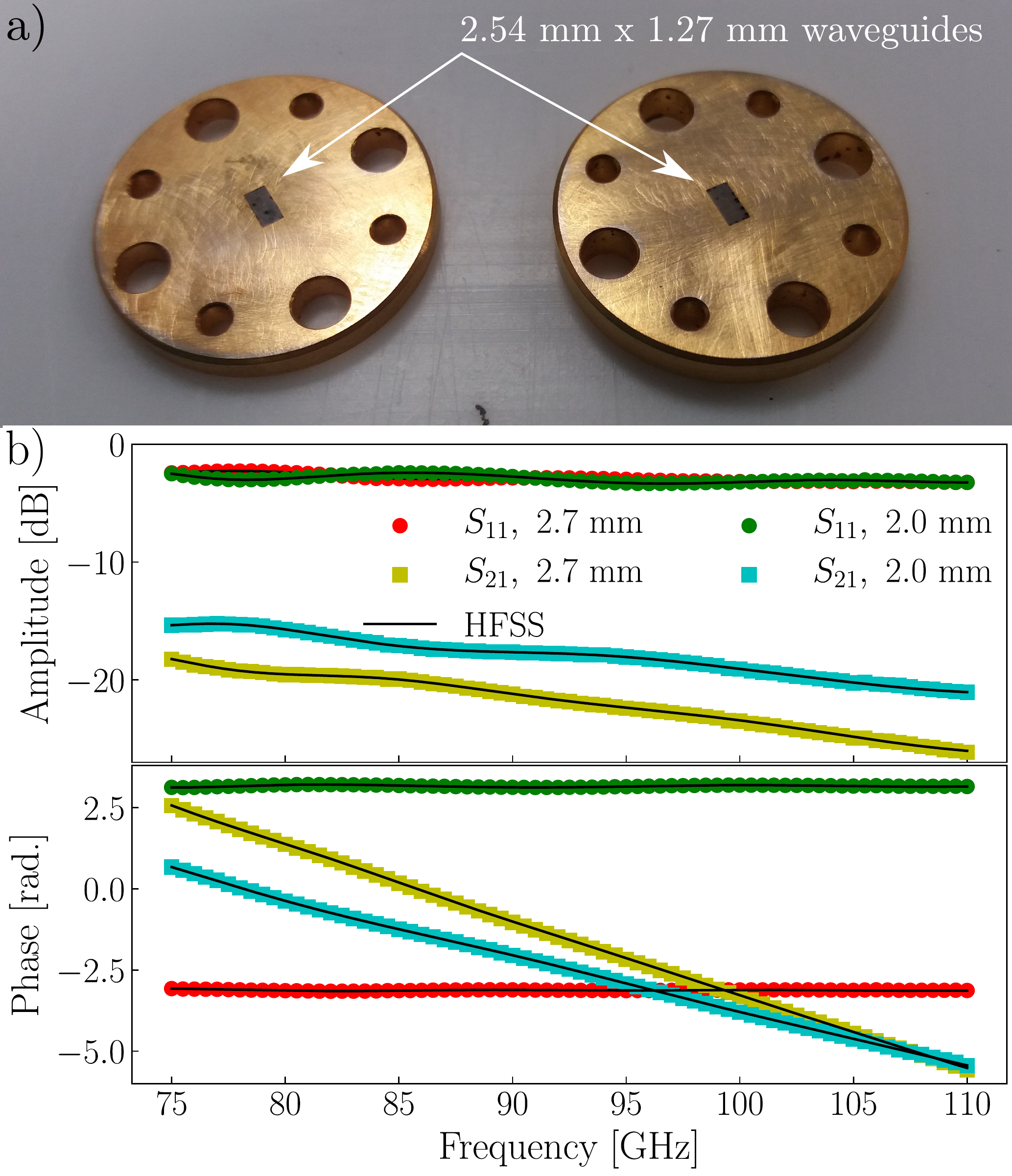}
	\end{minipage}
	\label{fig5}
\end{figure}

The setup is calibrated to measure the scattering parameters with respect to the reference planes at the facets of the waveguide spacers, and we can determine the relative permittivity $\epsilon_r(f)$ and permeability $\mu_r(f)$ of the Esorb-230 material using the Nicolson-Ross-Weir (NRW) method~\cite{NRW_1}.

\begin{figure}[h!]
	\begin{minipage}[t]{0.47\textwidth}
		\vspace{0mm}	
		\caption{Electromagnetic parameters of the Esorb-230 material in the frequency range from 75 to 110 GHz. (a) Real part of the relative dielectric permittivity. (b) Real part of the relative magnetic permeability. (c) Dielectric and magnetic loss tangents.}
	\end{minipage}
	\begin{minipage}[t]{0.47\textwidth}
		\vspace{1mm}
		\includegraphics[width=\textwidth]{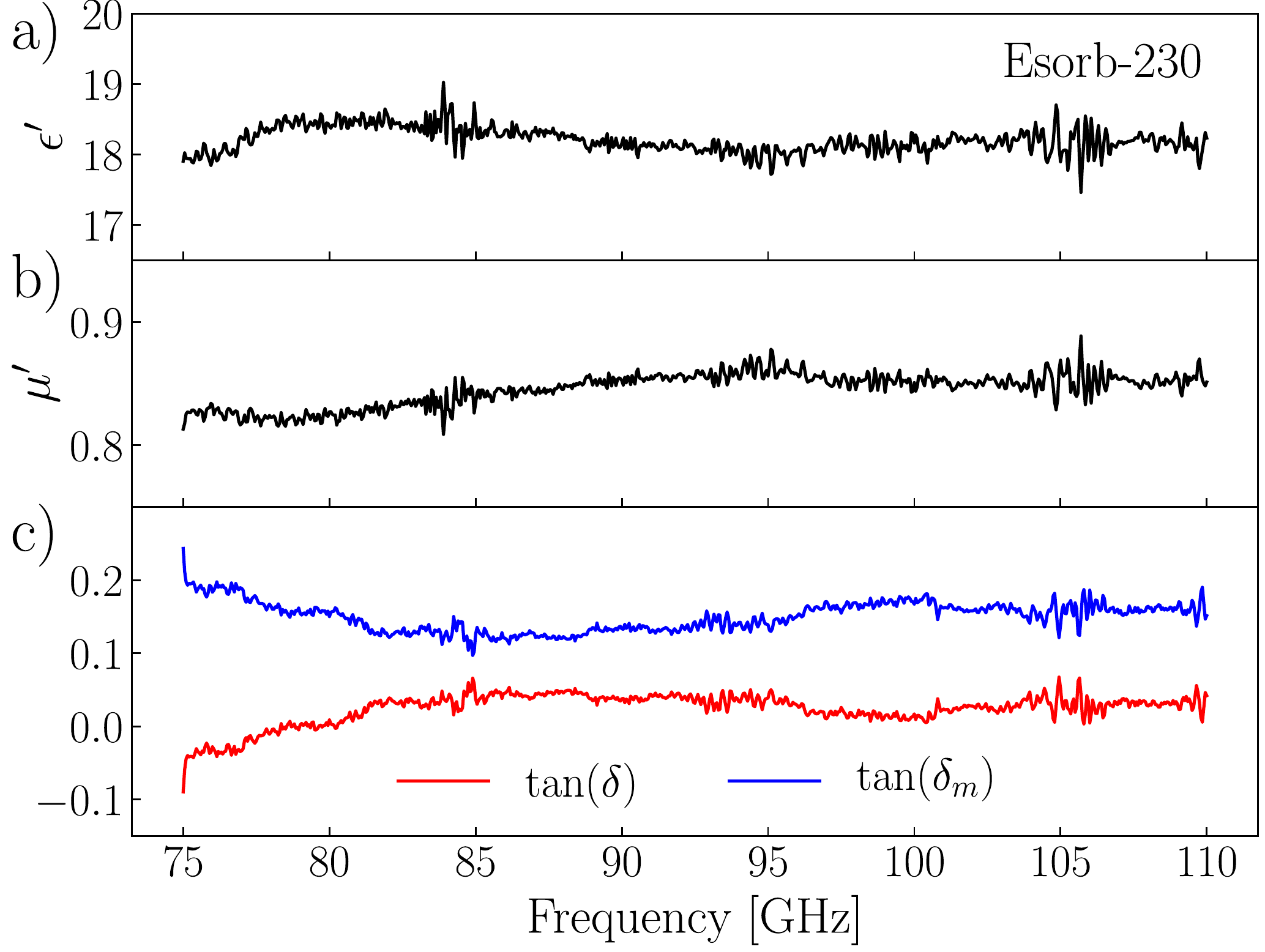}
	\end{minipage}
	\label{fig6}
\end{figure}

The method has ambiguity related to an {\it a priori} unknown branch of the complex logarithm function. Each branch gives a solution, and we enumerate the branches and corresponding solutions by index n. This ambiguity is resolved by having two sets of data obtained for two thicknesses of the waveguide sections (see Appendix 3). We determine the electromagnetic parameters $(\epsilon', \mu', \tan(\delta), \tan(\delta_m))$ for each section thickness and each branch $n$ and simulate the S-parameters with Ansys HFSS (High-Frequency Structure Simulator) for the experimental setting. We check the obtained electromagnetic parameters by comparing the simulated S-parameters with the measured ones and find a good agreement between them (Fig. 5(b)). The $\epsilon'(f)$ and $\mu'(f)$ found for the second branch of $d=2\ \rm{mm}$ thickness and the third branch of $d=2.7\ \rm{mm}$ thickness coincide well (see Appendix 3), and we determine all electromagnetic parameters as the mean values of these two solutions. The results are plotted in Fig.~\ref{fig6}.

\section{\label{sec:residual_photons}{Residual noise photons}}

We estimate the reduction in noise photon occupation number caused by the addition of the Esorb-230 filters at the mixing chamber plate of a dilution refrigerator. The number of photons reaching the filter input is described by the noise photon occupation number function computed earlier for the case with attenuators and length $l$ of the coaxial wiring (Fig.~\ref{motivation_figure}(b), blue line). For frequencies up to 70 GHz, where the filters were directly tested, we employ the measured $S_{21}$ parameters to estimate the upper bound of the number of photons transmitted through the filter. The filter attenuation $A(f)=P_{\rm{in}}(f)/P_{\rm{out}}(f)$ is
\begin{equation}
A=10^{-\frac{\overline{S_{21}}}{10}},\ \overline{S_{21}}=\left(\frac{1}{4}\sum_{i=1}^{4}(S_{21})_i\right)-(S_{21})_{\rm{thru}}.
\label{low_attenuation}
\end{equation}  
$\overline{S_{21}}$ is the mean value of the transmission coefficient over the four measured filters. The attenuation caused by the wiring is excluded by subtracting the transmission measured for the "through" connection $(S_{21})_{\rm{thru}}$.

\begin{figure}[h!]
	\begin{minipage}[t]{0.47\textwidth}
		\vspace{0mm}
		\caption{The estimation of reduction of noise photon occupation number caused by the use of the Esorb-230 filter or thinner UT047SS-SS and UT034SS-SS coaxial cables. The red line shows the case of the unfiltered UT086SS-SS coaxial line, where on average $\partial\overline{n}/\partial t\simeq 1420$ photons in the 82 to 110 GHz frequency range exit the end of the coaxial cable at the mixing chamber per second. When thinner cables are used, the average photon number is reduced (green and magenta lines). The use of the filter dramatically reduces the average flow of noise photons ($\partial\overline{n}/\partial t \ll 1$) for the same frequency range. Blue line is an estimate based on the measured electromagnetic properties of Esorb-230, and the black line is the upper bound estimated based on the measured S-parameters of the filter.}
	\end{minipage}
	\begin{minipage}[t]{0.47\textwidth}
		\vspace{1mm}
		\includegraphics[width=\textwidth]{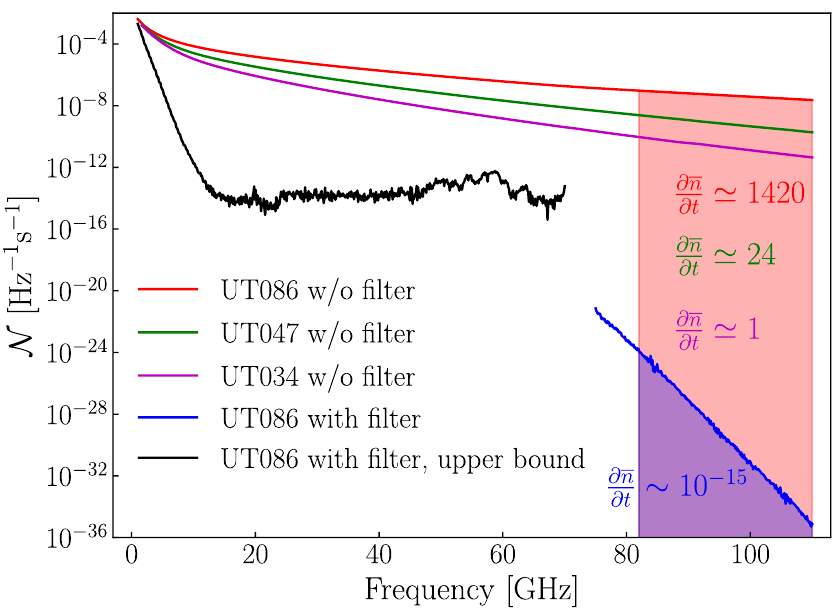}
	\end{minipage}
	\label{fig7}
\end{figure}

We compute the upper bound of the noise photon occupation number after the filter up to 70 GHz (Fig.~\ref{fig7}, black line) by using the filter attenuation from Eq.~(\ref{low_attenuation}) and Eq.~(\ref{attenuator_occupation}).

In the frequency range from 75 to 110 GHz, only the TEM and $\textrm{TE}_{11}$ modes are supported by the UT086SS-SS coaxial cable. Due to the reflection, the number of photons entering the filter per second, $\mathcal{N}_2$, is lower than the number of photons reaching the filter input per second, $\mathcal{N}_1$. This can be expressed as
\begin{equation}
\mathcal{N}_2 = \mathcal{N}_1\left(1-\left|\frac{Z_2-Z_1}{Z_2+Z_1}\right|^2\right).
\label{n_entrance}
\end{equation} 
For the TEM mode, $Z_1$ and $Z_2$ are the characteristic impedances of the coaxial line and the filter respectively (Eq.~(\ref{eq:impedance})). For the $\textrm{TE}_{11}$ mode, the characteristic impedance is not defined, and we use instead the wave impedances $Z_w(\omega)$ determined as
\begin{equation}
Z_w(\omega)=\frac{\omega\mu_0\mu_r(\omega)}{\sqrt{\frac{\omega^2}{c^2}\epsilon_r(\omega)\mu_r(\omega)-k_c^2}}.
\label{wave_impedance}
\end{equation} 
Here, $\epsilon_r$ and $\mu_r$ are the relative dielectric permittivity and magnetic permeability of the PTFE or Esorb material, and $k_c$ is the $\textrm{TE}_{11}$ mode critical wave-vector for the coaxial line or the filter.

Next, we compute the attenuation constants related to the conductor losses $\alpha_c$[dB/m] and the dielectric and magnetic losses $\alpha_{dm}$[dB/m] for both modes in the filter (see Appendix 4). Loss tangents of the Esorb-230 material are not small, and we compute the attenuation constants without assuming they are. We then use the simplified expressions to compute the attenuation as it was done earlier for coaxial lines and find only a small difference between the two ways. The losses in the conductors of the filter can be neglected as they are about three orders of magnitude smaller than the dielectric and magnetic losses in the Esorb-230 material. The total filter attenuation then reads
\begin{equation}
A(f)=10^{\frac{\alpha_{dm}(f)l}{10}},
\label{highr_attenuation}
\end{equation}
where $l=35.8$ mm is the length of the filter. This accounts for $\sim237\ \textrm{dB}$ at 100 GHz frequency for both modes. We use this frequency dependent attenuation (Eq.~(\ref{highr_attenuation})) and the number of photons entering the filter a second, $\mathcal{N}_2$, for each of the two modes in Eq.~(\ref{attenuator_occupation}) to get the number of noise photons transmitted through the filter per second. By summing up the results for two modes, we find an estimate for the noise photon occupation number, $\mathcal{N}\ \textrm{[Hz}^{-1}\textrm{s}^{-1}\textrm{]}$, transmitted through the filter in the frequency range from 75 to 110 GHz (Fig.~\ref{fig7}, blue line). 

The estimates of noise photon occupation number at the mixing chamber stage for thinner cables (UT047SS-SS and UT034SS-SS) of the same length and without the filter are shown for comparison in Fig.~\ref{fig7} (green and magenta lines). In these cases, the average number of noise photons per second reaching an experimental structure at millikelvin temperatures is reduced in the 82 to 110 GHz range - above the superconducting energy gap of the aluminium. Based on the material studies (Fig. 6) and the comparison of the room temperature and cryogenic microwave properties (Fig.3 and Fig.4), we expect the Esorb-230 filters to reduce the number of noise photons much more dramatically for the same frequency range, to values $\partial\overline{n}/\partial t\ll 1$. Despite the wiring with thinner and longer cables having an advantage in the noise photon flux, it cannot reduce the noise photon flux to the low values that are achievable with the Esorb-230 filters. Moreover, the thinner the cable, the stronger the dependence of attenuation on frequency, which makes the transmission steeper in the operational frequency band below 20 GHz. Though these estimates are made based on the room temperature measurements, the transmission properties of the filters will not change much at the cryogenic temperatures as can be seen in Fig.~\ref{fig:S_parameters} and Fig. 4.  

\section{\label{sec:conclusions}{Conclusions}}

The article provides an estimate of the spectral density of noise photons reaching an experimental structure at millikelvin temperatures per second through a coaxial line for frequencies up to 600 GHz. This elucidates the necessity of microwave to infrared filtering of coaxial wiring for experiments with quantum systems. The estimation is done for a dilution refrigerator wiring configuration typical for experiments with superconducting quantum circuits and UT086SS-SS, UT047SS-SS and UT034SS-SS cables. Cryogenic microwave frequency filters based on the CR-110 and Esorb-230 absorptive materials are manufactured. Transmission and reflection properties of the filters are tested up to 43 GHz cryogenically and up to 70 GHz at room temperature. Electromagnetic properties of the Esorb-230 material are separately measured in the frequency range from 75 to 110 GHz covering a range of frequencies above the superconducting gap of aluminium. Based on these measurements, the residual number of photons per second reaching a sample structure at millikelvin temperatures when the filters are used is estimated. The results help to construct coaxial wiring and filter solutions with reduced flux of noise photons of higher frequencies, including the pair breaking energies of metallic superconductors. 

\section{\label{ave_flow}{Appendix 1: Average flow of noise photons}}

To compute the noise photon occupation number $\mathcal{N}=\partial^2\overline{n}/\partial\nu\partial t$ -- the average number of noise photons in a unit bandwidth crossing a cross-section of microwave wiring line in a second -- we consider sections of coaxial line installed between the temperature stages of a dilution refrigerator from room temperature down to the mixing chamber plate. We assume a perfect thermalization at the temperature stages so that in equilibrium the temperature reduces linearly from the hot end, $T_H$, to the cold end, $T_C$,
\begin{equation}
T(x) = -\frac{(T_H-T_C)}{L}x+T_H.
\label{Eq1}
\end{equation} 
Here, $x$ denotes the position on the line from the hot end to the cold end, and L is the length of the coaxial line section connecting the two stages. 

If the coaxial cable has attenuation per unit length $\alpha(\omega)$[dB/m], a short section $dx$ of the line will have attenuation $\alpha dx$[dB] $= 10\log_{10}A$, where $A=P_{\rm in}/P_{\rm out}$ is the ratio of powers at the input and the output of the section. Noise photon occupation number at the input and the output of the section are related as
\begin{equation}
\mathcal{N}_{\rm out} = \frac{\mathcal{N}_{\rm in}}{A}+\frac{A-1}{A}n_{BE}\big(\omega,T(x)\big),
\label{Eq2}
\end{equation}
where $n_{BE}\big(\omega,T(x)\big)=1/(\exp[\hbar\omega/k_BT(x)]-1)$ is the Bose-Einstein distribution at frequency $\omega$ and temperature $T(x)$. Taking into account the smallness of $dx$, we can express through $\alpha$ the ratio $A=1+\frac{\alpha dx}{10}\ln 10$ and write Eq.~(\ref{Eq2}) as 
\begin{equation}
\mathcal{N}(x,\omega)+d\mathcal{N}=\frac{\mathcal{N}(x,\omega)}{1+\frac{\alpha dx}{10}\ln 10}+
\frac{\frac{\alpha dx}{10}\ln 10}{1+\frac{\alpha dx}{10}\ln 10}n_{BE}\big(\omega,T(x)\big).
\label{Eq3}
\end{equation}
Leaving only the terms up to the first-order in the small parameter $\alpha(\omega)dx$, we arrive at the first-order linear differential equation
\begin{equation}
\frac{\partial\mathcal{N}(x,\omega)}{\partial x}=\frac{\alpha(\omega)\ln 10}{10}\bigg(n_{BE}\big(\omega,T(x)\big)-\mathcal{N}(x,\omega)\bigg).
\end{equation}
Once the temperature distribution Eq.~(\ref{Eq1}) and the attenuation per unit length $\alpha$ are known, it is possible to solve this equation numerically and find the noise photon occupation number, $\mathcal{N}(x,\omega)$, as a function of the position $x$ along the coaxial line. The value $\mathcal{N}(0,\omega)$ here is known and serves as a boundary condition for the differential equation.

Starting from room temperature, where the boundary condition is $\mathcal{N}(0,\omega)=n_{BE}(\omega,T_{\rm RT})$, we can compute the function $\mathcal{N}(x,\omega)$ for the frequency $\omega$ consecutively for each section of coaxial line down to the mixing chamber plate. For those stages where attenuators are installed, the function $\mathcal{N}(x,\omega)$ will reduce abruptly according to Eq.~(\ref{Eq2}), with $\mathcal{N}_{\rm in}$ and $\mathcal{N}_{\rm out}$ being the function values before and after the attenuator and $n_{BE}$ the distribution corresponding to the temperature of the stage. The ratio $A=10^{a[dB]/10}$ here will be given by the attenuator value $a[dB]$. To perform the computation, we have to know the temperatures of the dilution refrigerator stages, the arrangement of attenuators at the temperature stages with their attenuations, and the lengths of coaxial line sections connecting each pair of stages.

We use in our computations temperatures $[300,\ 35,\ 2.85,\ 0.882,\ 0.082,\ 0.006]\ \rm K$ for room temperature, "50K", "4K", "Still Chamber", "Cold Plate", and "Mixing Chamber" stages, respectively. Lengths of the coaxial line sections are $[228,\ 271,\ 263,$ $\ 231,\ 306]\ \rm mm$ for interconnects between the neighbouring stages of the previous list, and the arrangement of attenuators is $[0,\ 20,\ 0,\ 20,\ 20]\ \rm dB$, meaning that $20\ \rm dB$ attenuators are installed at the "4K", "Cold Plate", and "Mixing Chamber" stages.

Integrating the function $\mathcal{N}(\omega,x_{\rm end})$ at the "Mixing Chamber" stage in a desired frequency range, we can find the average number of noise photons exiting the end of the line for this frequency range in a second.

\section{Appendix 2: Attenuation constants for different electromagnetic modes of a coaxial line}

We consider UT086SS-SS, UT047SS-SS and UT034SS-SS coaxial lines with stainless steel inner and outer conductors, PTFE dielectric, and the following parameters: electrical conductivity of stainless steel conductors $\sigma=1.41\cdot 10^6$ S/m, relative dielectric constant of the PTFE dielectric $\epsilon '=Re(\epsilon_r)=2.08$, relative permeability of PTFE $\mu '=Re(\mu_r)=1$, and the PTFE tangent of dielectric losses $\tan{\delta}=0.0004$. The radii of the inner conductors, $a$, and the inner radii of the outer conductors, $b$, are given in Table~\ref{coax_radii}. 

\begin{table}[h!]
	\caption{The radii of the inner conductors, $a$, and the inner radii of the outer conductors, $b$, for different coaxial lines used in the computation.}
	\begin{tabular}{||c||c|c|c||}
		\hline &UT086SS-SS&UT047SS-SS&UT034SS-SS\\
		\hline a [mm]&0.255&0.1435&0.1015\\
		\hline b [mm]&0.835&0.47&0.33\\
		\hline
	\end{tabular}
	\label{coax_radii}
\end{table}

The PTFE dielectric is not magnetic, and there are, therefore, only two types of energy losses for the coaxial lines -- dielectric losses and losses in conductors. Energy flow through a cross section of a coaxial line at the position $z$ along the line decreases as $P(z)=P_0\exp(-2\alpha z)$, where $\alpha$ is the attenuation constant and $P_0$ is the energy flow at the input of the line. In our case of non-magnetic materials, we let $\alpha=\alpha_c+\alpha_d$ to distinguish the two types of energy losses. We use cylindrical coordinates $(\vec{\rho},\vec{\phi},\vec{z})$ with the $z$-axis along the coaxial line direction. Taking the derivative of the above expression for the energy flow, we get $-P_l(z)=dP(z)/dz=-2\alpha P(z)$.  From here the attenuation constant can be expressed as
\begin{equation}
\alpha = \frac{P_l}{2P_0}.
\label{attenuation_conductor}
\end{equation}
$P_0$ here is the energy flow [W] at position zero, and $P_l$ is power dissipated per meter of line [W/m] at zero position.

The general expression to find $P_0$ is
\begin{equation}
P_0=\frac{1}{2}Re\int_{\rho=a}^{b}\int_{\phi=0}^{2\pi}(\vec{E}\times\vec{H}^*)_z\rho d\phi d\rho=\frac{1}{2}Re\int_{\rho=a}^{b}\int_{\phi=0}^{2\pi}(E_\rho H_\phi^*-E_\phi H_\rho^*)\rho d\phi d\rho.
\label{general_P_0}
\end{equation}
This expression is the integral of Umov-Poynting vector over the cross-sectional area of a coaxial line, and quantities under the integrals are electromagnetic field components.

Conductor losses per unit length can be found as
\begin{equation}
P_l = \frac{R_s}{2}\int_{C_1+C_2}\vec{H}_t\cdot\vec{H}_tdl=\frac{R_s}{2}\left[\int_{0}^{2\pi}(|H_z|^2+|H_\phi|^2)ad\phi+\int_{0}^{2\pi}(|H_z|^2+|H_\phi|^2)bd\phi\right].
\label{general_P_l}
\end{equation}
Here the integral is taken over the circumferences of the conductors, values of fields are taken at $\rho=a$ and $\rho=b$ for the first and the second integrals respectively, and $R_s$ is the surface resistance in [$\Omega$].

The attenuation constant related to the dielectric losses for TE or TM modes reads
\begin{equation}
\alpha_d = \frac{k^2\tan\delta}{2\beta}=\frac{k^2\tan\delta}{2\sqrt{k^2-k_c^2}},
\label{attenuation_dielectric}
\end{equation}
which is valid only if $k>k_c$. $k_c$ is a critical wave-vector.

The major mode for any coaxial transmission line is {\bf TEM mode} for which $E_z=H_z=0$, and the cut-off frequency is zero ($k_c=0$).

For the TEM mode, the attenuation constant $\alpha=\alpha_c+\alpha_d$ is equal to
\begin{equation}
\alpha_\textrm{TEM}= \frac{R_s}{4\pi Z_0}\left(\frac{1}{a}+\frac{1}{b}\right)+\frac{\pi\omega\epsilon_0\epsilon'\tan{\delta}Z_0}{\ln(b/a)}.
\label{TEM attenuation} 
\end{equation}

Here $R_s=\sqrt{\frac{\omega\mu_0\mu'}{2\sigma}}$ is the surface resistance in [$\Omega$] and $Z_0=\sqrt{\frac{\mu_0\mu'}{\epsilon_0\epsilon'}}\frac{\ln(b/a)}{2\pi}$ is the characteristic impedance of the coaxial line. The first term in Eq.~(\ref{TEM attenuation}) represents the attenuation related to losses in the conductors and the second term -- the attenuation related to the dielectric losses. This term can also be expressed as $\alpha_\textrm{d,TEM}=k\tan\delta/2$. Attenuation constants in Eqs.~(\ref{attenuation_conductor}), (\ref{attenuation_dielectric}) and (\ref{TEM attenuation}) are in units of Neper per meter [Np/m]. To convert these to [dB/m], one has to multiply the result by $20\log_{10}(e)$;  [dB/m]=$20\log_{10}(e)\cdot$[Np/m].

For {\bf TE modes} there is no electric field component along the coaxial line ($E_z=0$), and the cut-off frequency has a non-zero value. In this case, $\beta=\sqrt{k^2-k_c^2}$, where $k_c=2\pi f_c\sqrt{\epsilon'\mu'}/c>0$. All components of the field can be expressed through $H_z$ which, in turn, can be found from the Helmholtz equation and has a general form of 
\begin{equation}
H_z(\rho,\phi,z) = A\cos(n\phi)\left(CJ_n(k_c\rho)+DY_n(k_c\rho)\right)\exp(-j\beta z),
\label{TM_Hz}
\end{equation}
where $J_n(x)$ and $Y_n(x)$ are Bessel functions of the first and the second kind, respectively. Satisfying the boundary conditions on tangential components of electric field at conductor surfaces $E_\phi(\rho=a)=E_\phi(\rho=b)=0$, we arrive at the equation: $J_n^{'}(k_ca)Y_n^{'}(k_cb)-J_n^{'}(k_cb)Y_n^{'}(k_ca)=0$. The roots of the equation, $k_c^{n,m}$, where index $m$ denotes the $m$th root of the equation, are the cut-off wave-vectors determining the mode indexes $\textrm{TE}_{nm}$. Once the critical wave vectors $k_c$ are found, we can write down all components of the field
\begin{equation}
\begin{array}{l}
E_\rho=\frac{j\omega\mu_0\mu_r}{k_c^2\rho}An\sin(n\phi)\left(J_n(k_c\rho)-\frac{J_n^{'}(k_cb)}{Y_n^{'}(k_cb)}Y_n(k_c\rho)\right)\exp(-j\beta z),\\
E_\phi=\frac{j\omega\mu_0\mu_r}{k_c}A\cos(n\phi)\left(J_n^{'}(k_c\rho)-\frac{J_n^{'}(k_cb)}{Y_n^{'}(k_cb)}Y_n^{'}(k_c\rho)\right)\exp(-j\beta z),\\
H_\rho=\frac{-j\beta}{k_c}A\cos(n\phi)\left(J_n^{'}(k_c\rho)-\frac{J_n^{'}(k_cb)}{Y_n^{'}(k_cb)}Y_n^{'}(k_c\rho)\right)\exp(-j\beta z),\\
H_\phi=\frac{j\beta}{k_c^2\rho}An\sin(n\phi)\left(J_n(k_c\rho)-\frac{J_n^{'}(k_cb)}{Y_n^{'}(k_cb)}Y_n(k_c\rho)\right)\exp(-j\beta z),\\
H_z=A\cos(n\phi)\left(J_n(k_c\rho)-\frac{J_n^{'}(k_cb)}{Y_n^{'}(k_cb)}Y_n(k_c\rho)\right)\exp(-j\beta z).
\end{array}
\label{TE_field_components}
\end{equation}
In these equations the constant $A$ has units of [A/m].

Numerical calculations for UT086SS-SS, UT047SS-SS, and UT034SS-SS coaxial cables give us the TE mode's critical wave-vectors $k_c$ and cut-off frequencies $f_c$. The six TE modes with the lowest cut-off frequencies are given in Table~\ref{TE_table} for these three types of coaxial cables.

\begin{table}[h!]
	\caption{The six TE modes of UT086SS-SS, UT047SS-SS and UT034SS-SS coaxial cables with the lowest cut-off frequencies.}
	\begin{tabular}{|c||c|c||c|c||c|c||}
		\hline
		&\multicolumn{2}{c||}{UT086SS-SS}&\multicolumn{2}{c||}{UT047SS-SS}&\multicolumn{2}{c||}{UT034SS-SS}\\ \hline
		TE mode&$k_c$ [1/{\rm m}]&$f_c$ [GHz]&$k_c$ [1/{\rm m}]&$f_c$ [GHz]&$k_c$ [1/{\rm m}]&$f_c$ [GHz]\\ \hline11&1887&62.5&3352&111.0&4766&157.7\\ \hline21&3549&117.5&6305&208.7&8972&297.0\\
		\hline31&5004&165.6&8889&294.3&12658&419.1\\
		\hline01&5673&187.8&10077&333.6&14392&476.5\\
		\hline12&6176&204.5&10971&363.2&15652&518.2\\
		\hline41&6362&210.6&11303&374.2&16097&532.9\\
		\hline
	\end{tabular}
	\label{TE_table}
\end{table}

\begin{figure}[h!]
	\includegraphics[width=0.96\textwidth]{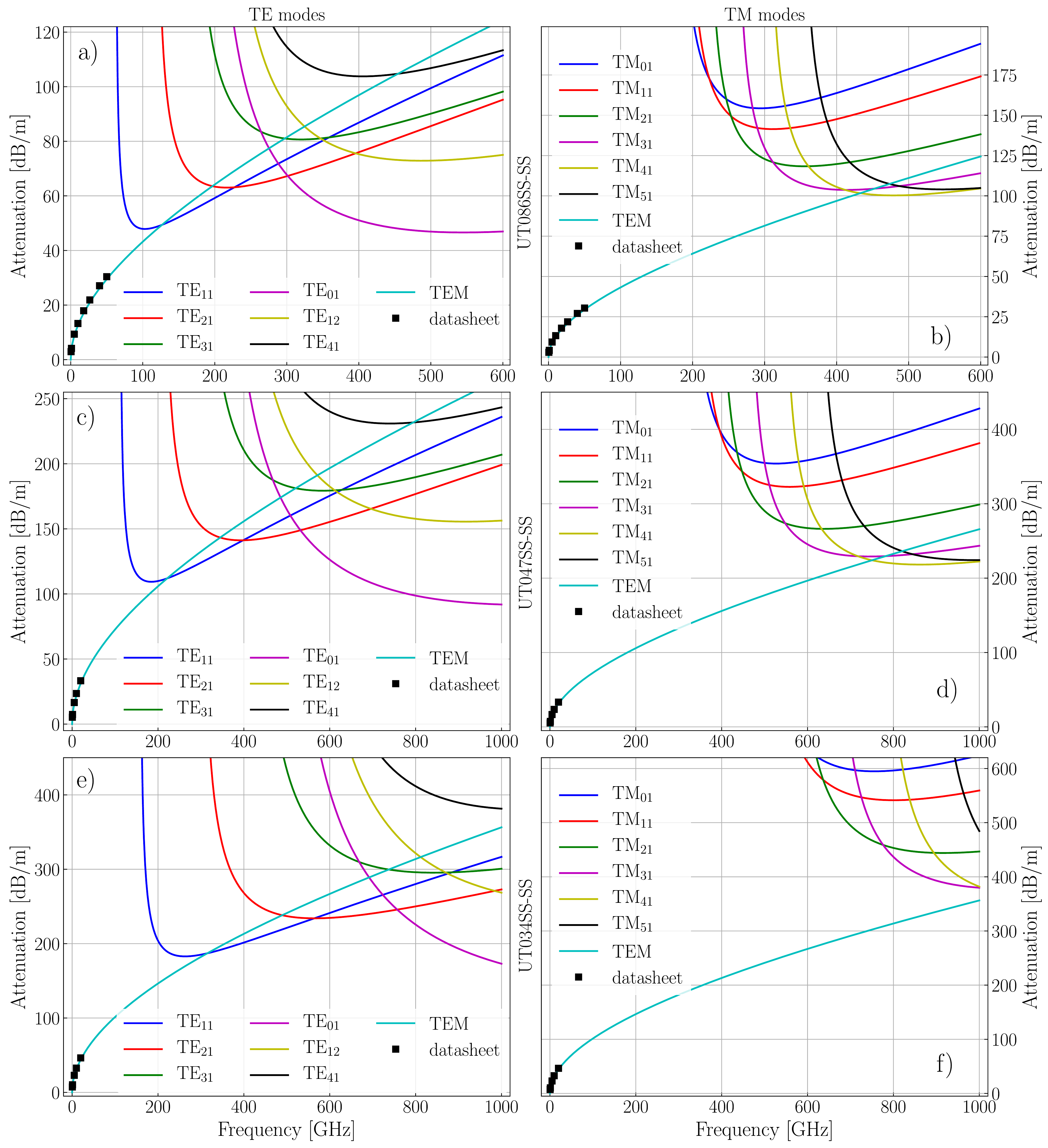}
	\caption{Attenuation per meter at room temperature for modes with the lowest cut-off frequencies: UT086SS-SS cable (a) TE, (b) TM modes, UT047SS-SS cable (c) TE, (d) TM modes, and UT34SS-SS cable (e) TE, (f) TM modes. Black squares represent the datasheet values for TEM modes.}
	\label{fig1}
\end{figure}

We find the energy flow at zero position, $P_0$, and the power dissipated per meter of line due to conductor losses, $P_l$, for a TE mode with the use of expressions (\ref{TE_field_components}) according to the formulas (\ref{general_P_0}) and (\ref{general_P_l})
\begin{equation}
\begin{array}{l}
P_0=\frac{\pi\omega\mu_0\mu'Re\{\beta\}|A|^2}{2k_c^4}\int\limits_{k_ca}^{k_cb}\left[\frac{n^2}{x}\left(J_n(x)-\frac{J_n^{'}(k_cb)}{Y_n^{'}(k_cb)}Y_n(x)\right)^2+x\left(J_n^{'}(x)-\frac{J_n^{'}(k_cb)}{Y_n^{'}(k_cb)}Y_n^{'}(x)\right)^2\right]dx,\quad{\rm and}\\
P_l=\frac{\pi|A|^2R_s}{2}\left\{a\left(1+\frac{|\beta|^2n^2}{k_c^4a^2}\right)\left[J_n(k_ca)-\frac{J_n^{'}(k_cb)}{Y_n^{'}(k_cb)}Y_n(k_ca)\right]^2+b\left(1+\frac{|\beta|^2n^2}{k_c^4b^2}\right)\left[J_n(k_cb)-\frac{J_n^{'}(k_cb)}{Y_n^{'}(k_cb)}Y_n(k_cb)\right]^2\right\}.
\end{array}
\label{TE_results}
\end{equation}
These quantities can be computed numerically, and we get $\alpha_{c,\textrm{TE}}$ according to Eq.~(\ref{attenuation_conductor}). Attenuation constants related to the dielectric losses are computed according to Eq.~(\ref{attenuation_dielectric}) for each mode with corresponding $k_c$ value. Total attenuation per meter at room temperature for the six TE modes with the lowest cut-off frequencies and TEM modes are shown in Fig.~\ref{fig1} for UT086SS-SS (a), UT047SS-SS (c), and UT034SS-SS (e) coaxial cable.

For {\bf TM modes}, there is no magnetic field component along the line ($H_z=0$). Modes of this type also have non-zero cut-off frequencies, $f_c$. This time, all components of the field can be expressed through the $E_z$ component. $E_z$ is a solution of the Helmholtz wave equation and has a general form which is the same as Eq.~(\ref{TM_Hz}). Components $E_z$ and $E_\phi$ have similar radial dependences, which includes $\left(CJ_n(k_c\rho)+DY_n(k_c\rho)\right)$. They both have to be zero at the surfaces of conductors to satisfy the boundary conditions which leads to the equation $J_n(k_ca)Y_n(k_cb)-J_n(k_cb)Y_n(k_ca)=0$. Importantly, here the Bessel functions enter the equation but their derivatives do not, in contrast to the TE modes. 

Numerical solutions of this equation for UT086SS-SS, UT047SS-SS, and UT034SS-SS coaxial cables give us the TM modes critical wave-vectors $k_c$ and cut-off frequencies $f_c$. The six TM modes with the lowest cut-off frequencies are given in Table~\ref{TM_table} for these types of coaxial cables.

\begin{table}[h!]
	\caption{The six TM modes of UT086SS-SS, UT047SS-SS, and UT034SS-SS coaxial cables with the lowest cut-off frequencies.}
	\begin{tabular}{|c||c|c||c|c||c|c||}
		\hline
		&\multicolumn{2}{c||}{UT086SS-SS}&\multicolumn{2}{c||}{UT047SS-SS}&\multicolumn{2}{c||}{UT034SS-SS}\\
		\hline TM mode&$k_c$ [1/\rm{m}]&$f_c$ [GHz]&$k_c$ [1/\rm{m}]&$f_c$ [GHz]&$k_c$ [1/\rm{m}]&$f_c$ [GHz]\\
		\hline01&5328&176.4&9464&313.3&13526&447.8\\
		\hline11&5673&187.8&10077&333.6&14392&476.5\\
		\hline21&6575&217.7&11680&386.7&16661&551.6\\
		\hline31&7789&257.9&13837.5&458.1&19721&652.8\\
		\hline41&9134&302.4&16228&537.2&23118&765.3\\
		\hline51&10518&348.2&18686.2&618.6&26616&881.2\\
		\hline
	\end{tabular}
	\label{TM_table}
\end{table}

Once the critical wave-vectors $k_c$ are found, we can determine all field components as
\begin{equation}
\begin{array}{l}
E_z=A\cos(n\phi)\left(J_n(k_c\rho)-\frac{J_n(k_cb)}{Y_n(k_cb)}Y_n(k_c\rho)\right)\exp(-j\beta z),\\
E_\rho=\frac{-j\beta}{k_c}A\cos(n\phi)\left(J_n^{'}(k_c\rho)-\frac{J_n(k_cb)}{Y_n(k_cb)}Y_n^{'}(k_c\rho)\right)\exp(-j\beta z),\\
E_\phi=\frac{j\beta}{k_c^2\rho}An\sin(n\phi)\left(J_n(k_c\rho)-\frac{J_n(k_cb)}{Y_n(k_cb)}Y_n(k_c\rho)\right)\exp(-j\beta z),\\
H_\rho=\frac{-j\omega\epsilon_0\epsilon_r}{k_c^2\rho}An\sin(n\phi)\left(J_n(k_c\rho)-\frac{J_n(k_cb)}{Y_n(k_cb)}Y_n(k_c\rho)\right)\exp(-j\beta z),\\
H_\phi=\frac{-j\omega\epsilon_0\epsilon_r}{k_c}A\cos(n\phi)\left(J_n^{'}(k_c\rho)-\frac{J_n(k_cb)}{Y_n(k_cb)}Y_n^{'}(k_c\rho)\right)\exp(-j\beta z).
\end{array}
\label{TM_field_components}
\end{equation}

With the known field components, we express the energy flow along the coaxial line, $P_0$, (Eq.~(\ref{general_P_0})) and the conductor losses per unit length, $P_l$, (Eq.~(\ref{general_P_l})) for TM modes, and the results read:
\begin{equation}
\begin{array}{l}
P_0=\frac{\pi\omega\epsilon_0\epsilon'Re\{\beta\}|A|^2}{2k_c^4}\int\limits_{k_ca}^{k_cb}\left[x\left(J_n^{'}(x)-\frac{J_n(k_cb)}{Y_n(k_cb)}Y_n^{'}(x)\right)^2+\frac{n^2}{x}\left(J_n(x)-\frac{J_n(k_cb)}{Y_n(k_cb)}Y_n(x)\right)^2\right]dx,\\
P_l=\frac{\pi R_s\omega^2\epsilon_0^2\epsilon'^2|A|^2}{2k_c^2}\left[a\left(J_n^{'}(k_ca)-\frac{J_n(k_cb)}{Y_n(k_cb)}Y_n^{'}(k_ca)\right)^2+b\left(J_n^{'}(k_cb)-\frac{J_n(k_cb)}{Y_n(k_cb)}Y_n^{'}(k_cb)\right)^2\right].
\end{array}
\label{TM_results}
\end{equation}
Constant $A$ has units of [V/m] in these equations. These quantities can be computed numerically for different TM modes, and we get $\alpha_{c,\textrm{TM}}$ according to Eq.~(\ref{attenuation_conductor}). Attenuation constants related to the dielectric losses for TM modes are computed in the same way as for the TE modes in accordance with Eq.~(\ref{attenuation_dielectric}) with critical wave-vectors $k_c$ obtained for the TM modes. Total attenuation per meter at room temperature for the six TM modes with the lowest cut-off frequencies and the TEM modes are shown in Fig.~\ref{fig1} for (b) UT086SS-SS, (d) UT047SS-SS, and (f) UT034SS-SS coaxial cables. As can be seen from Fig.~\ref{fig1}, the attenuation of TM modes is considerably higher (by at least 50 dB) than that of TE modes for these types of cables. The contribution of the TM modes to the total flux of noise photons is negligible, and we use only attenuation of the TE modes in the estimates of the noise photon flux. 

\begin{figure}[h!]
	\includegraphics[width=0.96\textwidth]{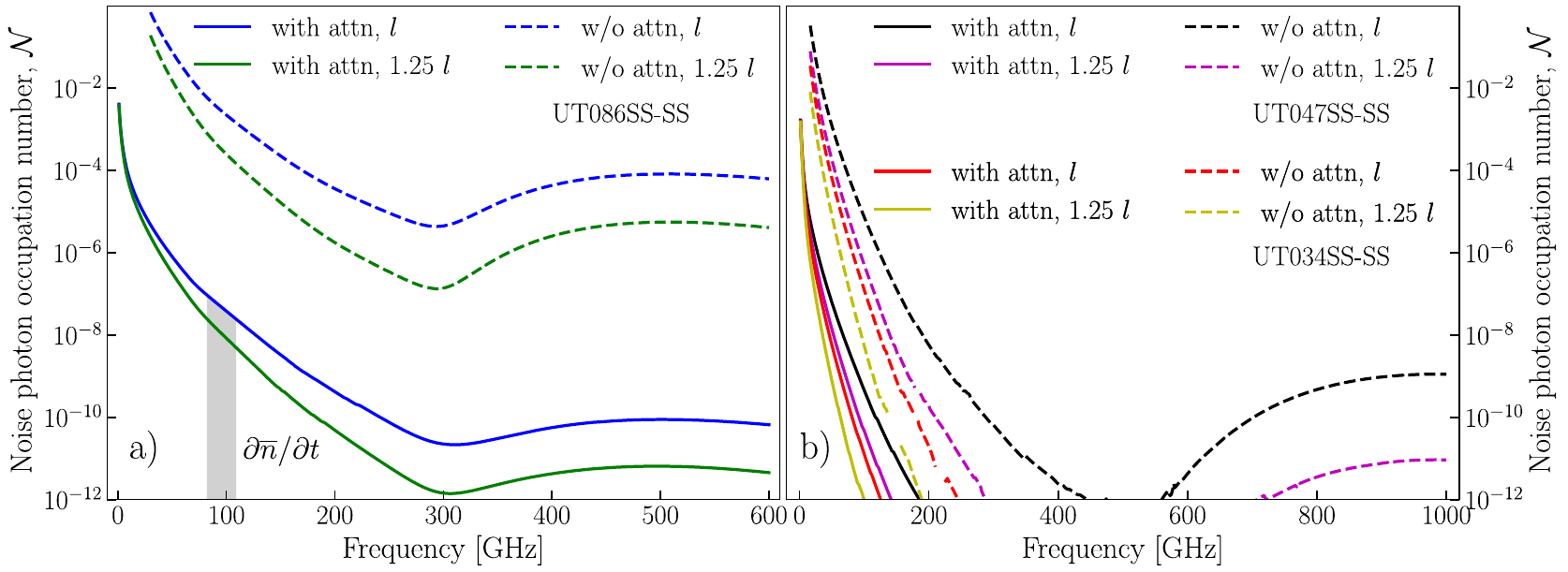}
	\caption{Estimates of the noise photon occupation numbers for the UT086SS-SS coaxial cable (a) and UT047SS-SS and UT034SS-SS coaxial cables (b). The estimation does not take into accout the temperature dependence of the attenuation, and the lines represent the lower bounds of the noise photon occupation numbers.}
	\label{fig_s2}
\end{figure}

When the total attenuation is known for a given mode, it is possible to estimate the noise photon occupation number for this particular mode at the frequencies of interest as was explained in Appendix 1. Once these numbers are computed for all modes, we can sum up all of them to get the total noise photon occupation number, $\mathcal{N}(\omega,x_{\rm end})$. Fig.~\ref{fig_s2} shows the estimates of the noise photon occupation numbers at the mixing chamber stage of a dilution refrigerator for UT086SS-SS cable (a) and UT047SS-SS and UT034SS-SS cables (b). The thinner the cable, the higher the attenuation for the same type of mode. As a result, the estimates of the noise photon occupation number are noticeably lower for thinner coaxial cables. The estimates are made for the wiring configuration described in Appendix 1. Reduction in the cables attenuation for lower temperatures is not taken into account, and the actual noise photon occupation numbers will be higher than those shown in Fig.~\ref{fig_s2}. The lines in Fig.~\ref{fig_s2} represent the lower bounds of the noise photon occupation numbers. The thinner the cable, the bigger the deviation of the actual noise photon occupation number from the presented values. 

\section{Appendix 3: Determination of electromagnetic properties of Esorb-230 material} 

\begin{figure}[h!]
	\includegraphics[width=0.96\textwidth]{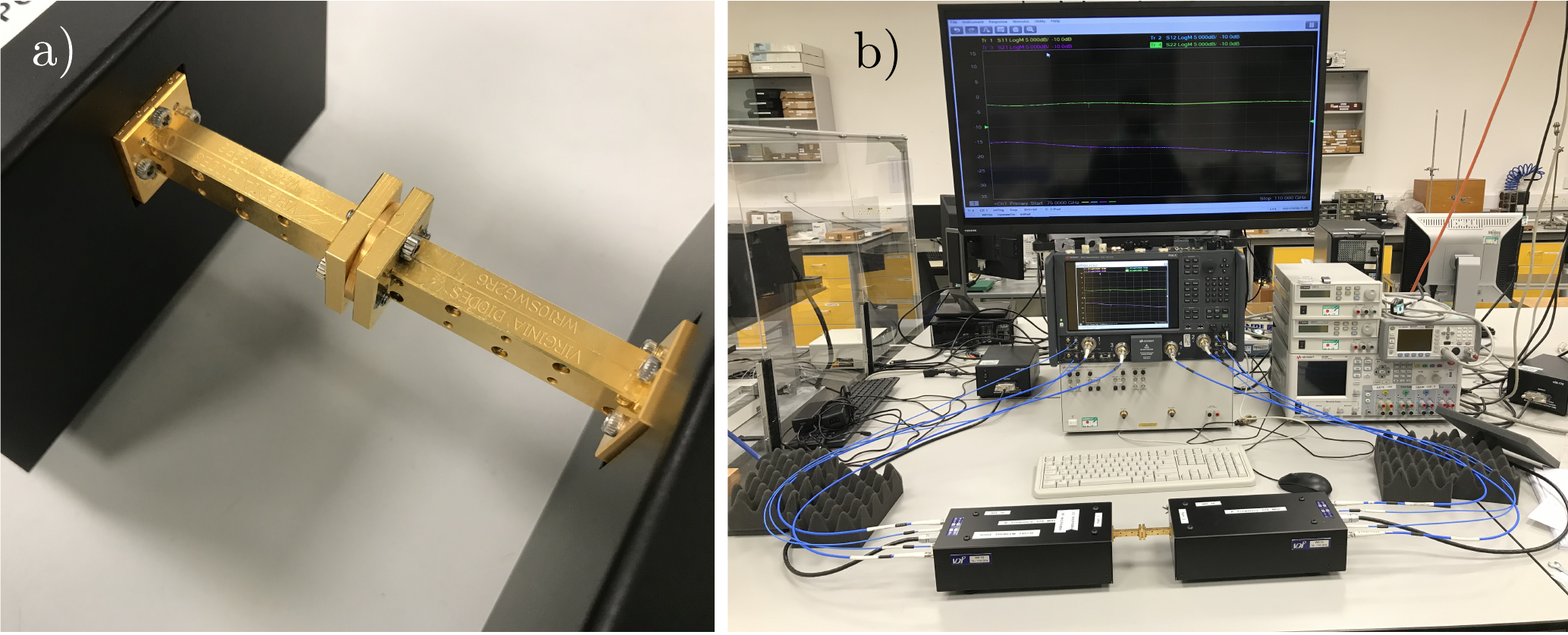}
	\caption{Measurements of $S_{11}$ and $S_{21}$ scattering parameters of rectangular waveguide sections filled with the Esorb-230 material. (a) The waveguide section placed between the fixtures of the microwave setup. (b) The microwave setup used in the measurements.}
	\label{fig_s3}
\end{figure}

Measurements of $S_{11}$ and $S_{21}$ scattering parameters were performed on two thin sections of rectangular waveguides filled with the Esorb-230 absorptive material. Two WR10 waveguides (2.54 mm $\times$ 1.27 mm) with thicknesses 2 mm and 2.7 mm were used in the measurements. Figure~\ref{fig_s3}(a) shows the waveguide section placed between two fixtures of the microwave setup frequency extension modules. The microwave setup operates in the frequency range from 75 to 110 GHz and consists of a VNA with frequency extension modules is shown in Fig.~\ref{fig_s3}(b). 
The system is calibrated to measure with respect to reference planes at the both sides of the waveguide sections. The measured $S_{11}(f)$ and $S_{21}(f)$ parameters are shown in Fig. 5(b) of the main text. 

According to the Nicolson-Ross-Weir method, the complex relative permittivity $\epsilon_r$ and permeability $\mu_r$ for our experimental setting can be found as
\begin{equation} 
\mu_r(f)=\frac{k_z}{\sqrt{(\frac{\omega}{c})^2-(\frac{\pi}{a})^2}F(f)},\quad\epsilon_r(f)=\frac{c^2\Big(k_z^2+(\frac{\pi}{a})^2\Big)}{\omega^2\mu_r(f)}, 
\label{eps_mu}
\end{equation}
where $k_z$ is the component of complex wave-vector along the direction of the waveguide, $a=2.54$~mm is the bigger side of the rectangular waveguide, and $F(f)=(1-\Gamma(f))/(1+\Gamma(f))$. Interfacial reflection coefficient $\Gamma(f)$ is defined as
\begin{equation}
\Gamma(f)= \frac{1-S_{21}^2+S_{11}^2}{2S_{11}}+\sqrt{\left(\frac{1-S_{21}^2+S_{11}^2}{2S_{11}}\right)^2-1}.
\label{refl_coeff}
\end{equation}   
To compute $k_z$, we first obtain the propagation factor $P(f)$ as
\begin{equation}
P(f) = \vert P(f)\vert e^{j\phi(f)} = \frac{S_{21}(f)-S_{11}(f)+\Gamma(f)}{1+\Gamma(f)(S_{21}(f)-S_{11}(f))}=e^{-jk_zd},
\label{P_factor}
\end{equation}
where d is the thickness of the waveguide section filled with the absorptive material. From here,
\begin{equation}
Re(k_z) = \frac{2\pi n-\phi(f)}{d},\quad\textrm{and}\quad Im(k_z) = \frac{\ln\vert P(f)\vert}{d}.
\label{re_im_kz}
\end{equation}
The phase of the propagation factor is $2\pi$ periodic and, as a result, the real part of the wave-vector depends on the unknown integer, $n$, enumerating the solution branch. This is where the ambiguity of the method arises. 

The dielectric and magnetic loss tangents are found from $\epsilon_r$ and $\mu_r$ as
\begin{equation}
\tan(\delta) = -\frac{Im(\epsilon_r)}{Re(\epsilon_r)},\quad\rm{and}\quad\tan(\delta_m)=-\frac{Im(\mu_r)}{Re(\mu_r)}.
\label{td_tdm}
\end{equation} 

\begin{figure}[h!]
	\includegraphics[width=0.96\textwidth]{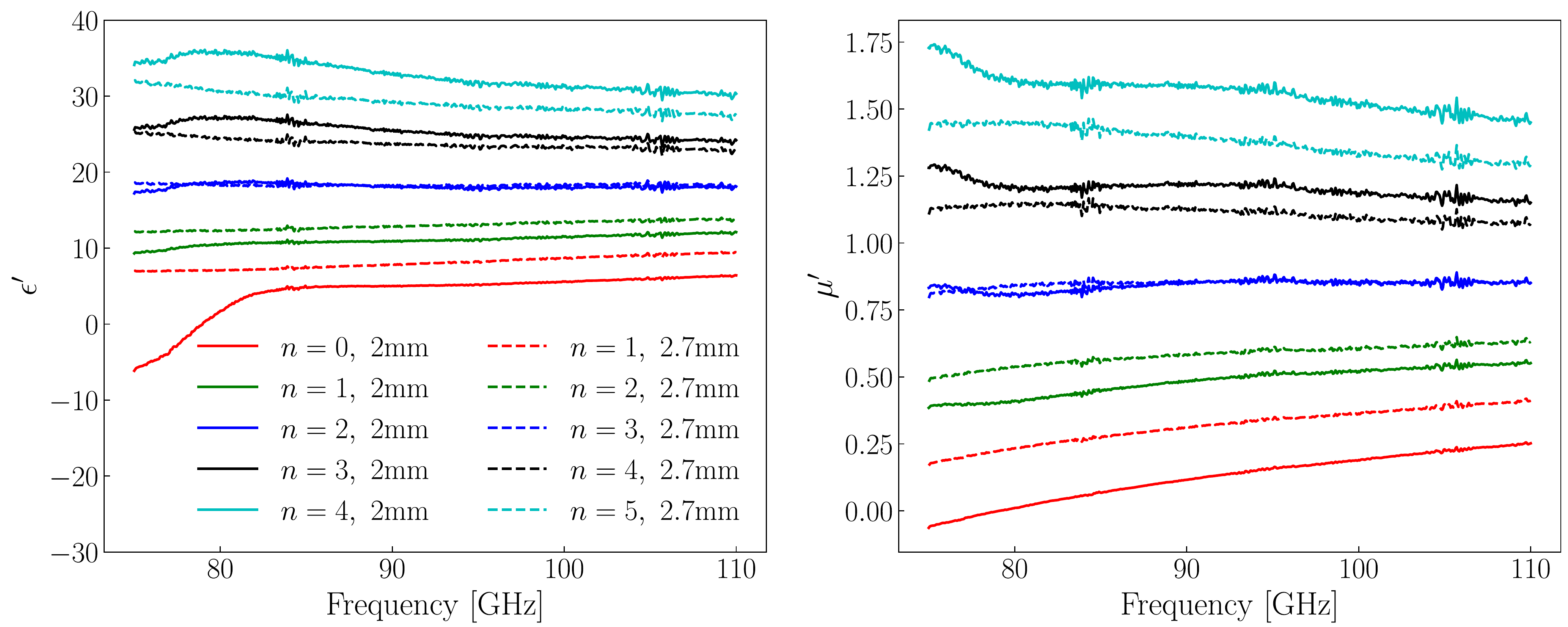}
	\caption{Real parts of relative dielectric permittivity $\epsilon^\prime$ and magnetic permeability $\mu^\prime$ of the Esorb-230 material in the frequency range from 75 to 110 GHz determined by the NRW method. Data for two waveguide section thicknesses $d=2.0\ \rm{mm}$ and $d=2.7\ \rm{mm}$ and different solution branches are shown.}
	\label{fig_s4}
\end{figure}

We calculate the real parts of dielectric permittivity $\epsilon'=Re(\epsilon_r)$ and magnetic permeability $\mu'=Re(\mu_r)$ for both waveguide sections with thicknesses $d=2.0\ \rm{mm}$ and $d=2.7\ \rm{mm}$ and different branch indexes $n$, see Fig.~\ref{fig_s4}. Electromagnetic parameters should not depend on the section thickness, and this allows for the resolution of the ambiguity. Namely, the real parts $\epsilon'$ and $\mu'$ determined for $n=2$ and $d=2.0\ \rm{mm}$ and for $n=3$ and $d=2.7\ \rm{mm}$ closely coincide, in contrast to the solutions for the other branches. The loss tangents given by the imaginary parts of $\epsilon_r$ and $\mu_r$ are much smaller, and the branch selection is made based on the real parts of the quantities. The final values of all electromagnetic parameters of Esorb-230 material ($\epsilon',\ \mu',\ \tan(\delta),\ \tan(\delta_m)$) are taken as mean values of the solutions for these two branches.

\section{Appendix 4: Attenuation constant of the Esorb-230 filter}

Only two electromagnetic modes are supported by the UT086SS-SS coaxial line in the frequency range from 75 to 110 GHz. These are the TEM and $\textrm{TE}_{11}$ modes. With the knowledge of the electromagnetic parameters of the Esorb-230 material and the geometric parameters of the filter, it is possible to estimate the attenuation constants for these two modes using a similar method to the one shown earlier for the coaxial line. We compute the critical wave-vector $k_c=645\ 1/$m and the cut-off frequency $f_c=8.1$ GHz for the $\textrm{TE}_{11}$ mode. The values are reduced in comparison to the coaxial line due to the bigger diameters of the inner and outer conductors of the filter and the bigger relative dielectric constant of the absorptive material.

\begin{figure}[h!]
	\includegraphics[width=0.96\textwidth]{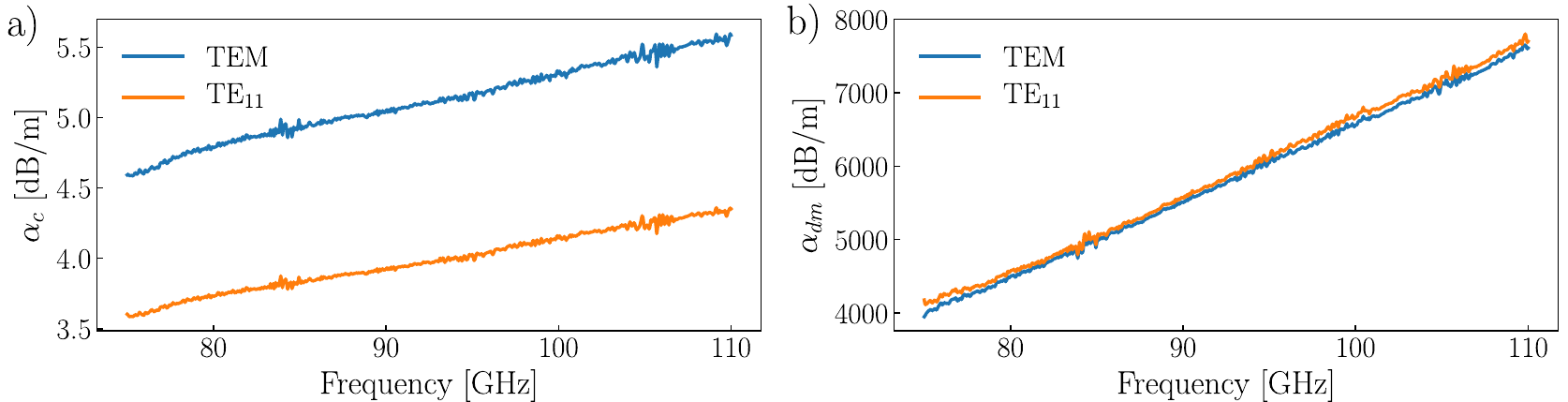}
	\caption{Attenuation constants of the filter for TEM and TE$_{11}$ modes due to (a) conductor losses and (b) dielectric and magnetic losses in the Esorb-230 material.}
	\label{fig_s5}
\end{figure}

With the known electromagnetic fields inside the filter, we can compute the integrals $P_0$ Eq.~(\ref{general_P_0}) and $P_l$ Eq.~(\ref{general_P_l}) for both modes to estimate the attenuation related to the conductor losses per unit length of the filter $\alpha_c\  [\textrm{dB/m}] = 20\log_{10}(e)\cdot \big(P_l/2P_0\big)$. The room temperature value of copper conductivity, $\sigma_{\textrm{Cu,RT}}\simeq 67\cdot10^6$ S/m, was used in the surface resistance estimation. The obtained results are shown in Fig.~\ref{fig_s5}(a). Loss tangents of Esorb-230 material are not small, and we find the dielectric $P_d$ and magnetic $P_m$ losses per unit length of the filter by computing the integrals 
\begin{equation}
P_d = \frac{\omega}{2}\int_{\rho=a}^{b}\int_{0}^{2\pi}\epsilon''|\vec{E}|^2\rho d\phi d\rho\quad\textrm{and}\quad
P_m = \frac{\omega}{2}\int_{\rho=a}^{b}\int_{0}^{2\pi}\mu''|\vec{H}|^2\rho d\phi d\rho.
\label{em_losses}
\end{equation}
Here, $\epsilon''=\epsilon_r \tan(\delta)$ and $\mu''=\mu_r\tan(\delta_m)$ are the imaginary parts of the dielectric permittivity and the magnetic permeability. The attenuation related to the material losses per unit length is found as $\alpha_{dm}\ [\textrm{dB/m}] = 20\log_{10}(e)\cdot \big((P_d+P_m)/2P_0\big)$. The results are shown in Fig.~\ref{fig_s5}(b). 

Attenuation constants given in Eq.~(\ref{TEM attenuation}) and Eq.~(\ref{attenuation_dielectric}) are good approximations for the case when losses in the insulating material of a coaxial line are small. We find that the attenuation constants computed with the use of these equations deviate only a little bit from the constants presented in Fig.~\ref{fig_s5} for the entire studied frequency range. The deviation does not exceed $3.7\%$ of the values given in Fig. 12.

Conductor losses are about three orders of magnitude lower than the dielectric and magnetic losses and can be neglected in the consideration of filter attenuation. Dielectric and magnetic losses for TEM and $\textrm{TE}_{11}$ modes are very similar due to the small value of the critical wave-vector $k_c$ for the $\textrm{TE}_{11}$ mode in comparison to the wave-vectors for the considered frequency range.


\begin{backmatter}


\section{Funding}
The authors are thankful for the support from the  European Research  Council  (ERC)  under  the  Grant  Agreement No.  648011, the Engineering and Physical Sciences Research Council (EPSRC) under the Grant Agreements No. EP/T018984/1 and No. EP/T001062/1, the Scottish Research Partnership in Engineering (SRPe) under the Grant Agreement NMIS-IDP/025 and Seeqc UK LIMITED for a studentship (JB).


\section{Availability of data and materials}
The datasets used and/or analysed during the current study are available from the corresponding author on reasonable request.


\section{Competing interests}
The authors declare that they have no competing interests.


\section{Authors' contributions}
JBa, ZZ, and SD took the microwave measurements. MF made HFSS simulations, XS and NR took the scattering parameter measurements in the extremely high frequency band. JBu helped with the manufacturing of filters. NR, CL, and MW supervised the work. SD estimated the noise photon flux, designed the filters, processed the data, and wrote the manuscript with contributions from other authors.

\section{Acknowledgements}
The authors are thankful to Nicholas Nugent for careful reading and checking the manuscript language.



\bibliographystyle{vancouver} 
\bibliography{Manuscript}      








%

\end{backmatter}
\end{document}